\documentclass[onecolumn]{aastex631}

\usepackage{natbib}
\usepackage{savesym}
\savesymbol{tablenum}
\usepackage{siunitx}
\restoresymbol{SIX}{tablenum}

\shorttitle{TONGS: A Treasury of Nearby Galaxy Surveys}
\shortauthors{Christie et al.}

\graphicspath{{./}{figures/}}

\begin{document}

\title{TONGS: A Treasury Of Nearby Galaxy Surveys}

\author[0009-0004-8163-6293]{Hannah S. Christie}
\affiliation{Department of Physics \& Astronomy and Institute
for Earth and Space Exploration\\
University of Western Ontario \\
1151 Richmond St.\\
London ON, Canada}

\author[0000-0002-2620-6483]{Adrien Hélias}
\affiliation{Department of Physics \& Astronomy\\
University of Western Ontario \\
1151 Richmond St.\\
London ON, Canada}

\author[0000-0002-5568-6965]{Matheus do Carmo Carvalho}
\affiliation{Department of Physics \& Astronomy\\
University of Western Ontario \\
1151 Richmond St.\\
London ON, Canada}

\author[0000-0003-2767-0090]{Pauline Barmby}
\affiliation{Department of Physics \& Astronomy and Institute
for Earth and Space Exploration\\
University of Western Ontario \\
1151 Richmond St.\\
London ON, Canada}

\begin{abstract}
The beginning of the 21st century marked the ``modern era of galaxy surveys" in astronomy. Rapid innovation in observing technology, combined with the base built by galaxy catalogs and atlases dating back centuries, sparked an explosion of new observational programs driven by efforts to understand the different processes driving galaxy evolution. This review aims to answer the following science questions: (1) how have galaxy surveys evolved in the past 20 years, and how have traditional observational programs been affected by the rise of large panoramic surveys, (2) can the term ``nearby" be quantified in the context of galaxy surveys, and (3) how complete is the coverage of the nearby universe and what areas hold the largest opportunity for future work? We define a galaxy survey as a systematically obtained data set which aims to characterize a set of astronomical objects. Galaxy surveys can further be subdivided based on the methods used to select the objects to observe, the properties of the survey samples (e.g. distance or morphology), or the observing strategies used. We focus on \textit{pointed} nearby galaxy surveys, which we define as surveys which observe a specific sample of target galaxies. Through a study of 43 nearby galaxy surveys, we find no standardized quantitative definition for ``nearby" with surveys covering a wide range of distances. We observe that since 2003, traditional targeted galaxy surveys have undergone a dramatic evolution, transitioning from large, statistical surveys to small, ultra-specific projects which compliment the rise of large high resolution panoramic surveys. While wavelength regimes observable from the ground (such as radio or optical wavelengths) host numerous surveys, the largest opportunity for future work is within the less covered space-based wavelength regimes (especially ultraviolet and X-ray). 
\end{abstract}


\section{Introduction} \label{sec:intro}
The recent ``data-flooding" of astronomy has included an explosion of observational galaxy surveys, with the beginning of the 21st century being referred to as the ``modern era for galaxy surveys" \citep{okamura2020}. The rapid growth of the field is matched only by the development of new observing technology. The large number of telescopes and satellites launched in the past 20 years have not only increased the overall coverage of the nearby galaxies in terms of wavelength completeness, but also expanded the parameter space possible to observe \citep{djorgovski2013}. More sensitive observing technology has allowed for the spatially resolved study of fainter, smaller, and more distant galaxies. 

Observational studies have identified a largely bimodal distribution of galaxies in which young galaxies begin their lifespans blue in color, gas rich, and star forming. Then, over vast timescales, longer than possible to observe, they begin to transition to quiescence where they are observed to be redder in color, gas poor and lacking in new stars \citep{Strateva2021}. To further complicate the picture, there are observed populations of galaxies located in the ``green valley" that display a mixed set of properties \citep{cleland2021}. While many galaxies appear on a global scale to exist in a similar stage of evolution, a galaxy's individual properties may vary drastically depending on the type of galaxy, age, or environment. Galaxy surveys compile large numbers of different observations in order to create a more complete understanding of these processes. These surveys can be divided into smaller type groups: panoramic sky surveys or pointed galaxy surveys. For the purposes of this paper, we will focus on \emph{pointed} surveys of \emph{nearby} galaxies, and the effects of new wide-coverage sky surveys on traditional survey techniques.
In this paper, we aim to answer the following questions:

\begin{enumerate}
  \item What constitutes galaxy surveys, and why are they important?
  \item What does \emph{nearby} mean in terms of distances to targets?
  \item What is the current coverage from pointed galaxy surveys?
  \item How have nearby galaxy surveys been affected by the rise of large panoramic observing programs?
\end{enumerate}

To answer these questions, we reviewed a sample of nearly 50 nearby galaxy surveys. Nearby galaxy surveys take advantage of the small distances to targets which allow for high resolution studies to be performed. This increases the types of parameters that can be measured, giving a deeper understanding of each galaxy studied. The numerous galaxies located within the Local Volume also provide an ideal laboratory to learn more about our own galaxy and its evolution. In an attempt to avoid redundancies, we have opted to exclude surveys which have been superseded with the release of new data sets or survey missions. We further only include surveys of individual galaxies, rather than galaxy clusters or groups, to focus our scope. Our goal in this paper is to summarize the survey samples and observations, rather than their scientific results, for which we refer the reader to the survey teams' paper. 

This paper is organized as follows: in Section \ref{sec: Overview}  we  provide a brief overview of galaxy surveys and their recent evolution, Section \ref{sec: Selection} discusses the selection criteria for the sample of nearby galaxy surveys discussed in this paper. Sections \ref{sec: radio}-\ref{sec: xray} contain brief summaries for the sample of surveys (sorted by wavelength and chronologically within wavelength sections), and Section \ref{sec: Discussion} discusses the above questions, as well as an overview of all nearby galaxy surveys included in this paper. 

\begin{figure} [ht!]
\centering
\includegraphics[width=.8\textwidth]{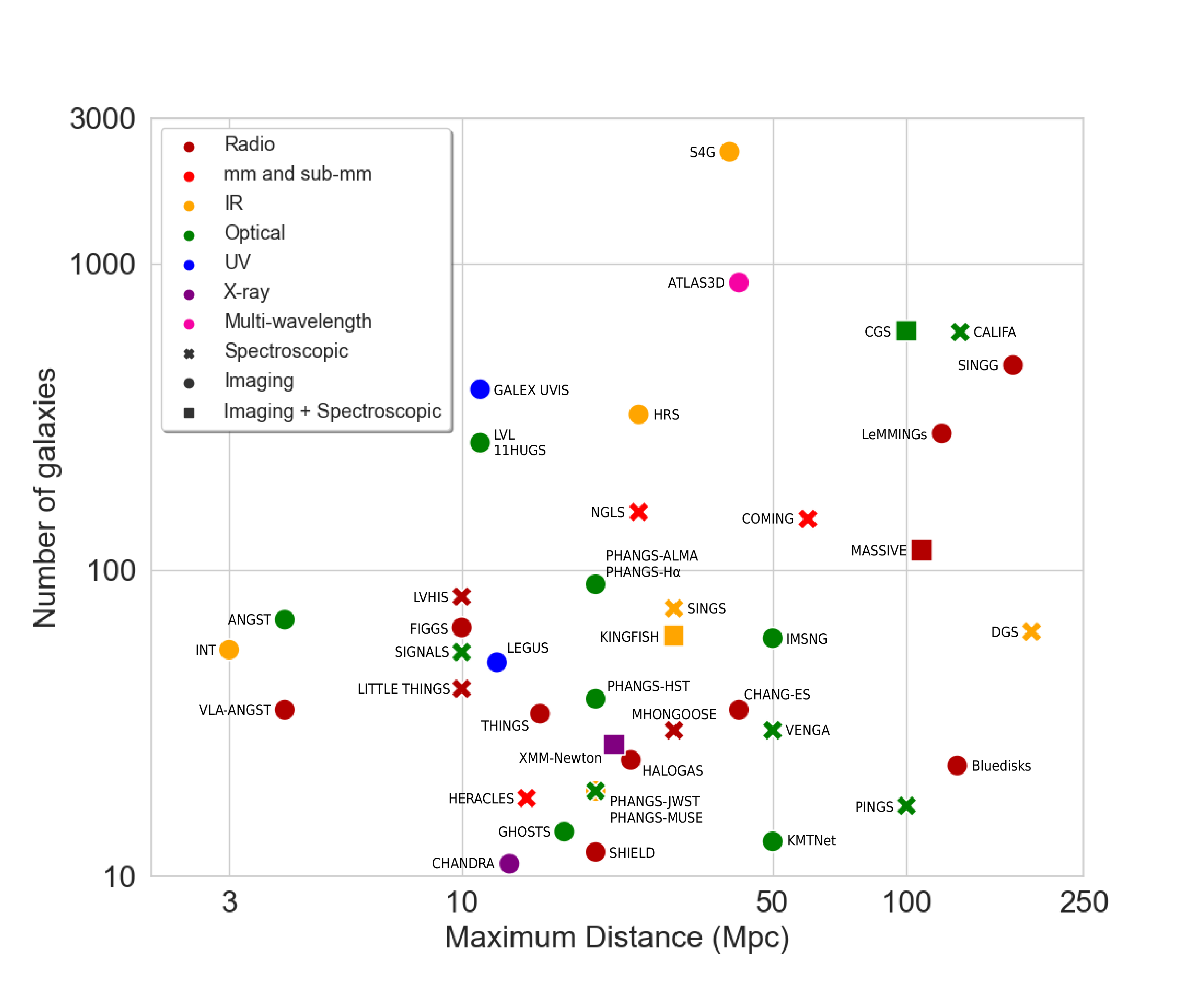}
\caption{Number of galaxies and maximum distance in Mpc for each of the surveys cited in this paper. The color represents the wavelength domain probed. The shape of the markers represent the type of data collected (imaging, spectroscopy, or both). UV surveys have the most limited distance range while X-ray surveys display the smallest sample sizes. Radio and Optical surveys span the widest range in terms of both sample size and distance coverage.}
\label{fig:1}
\end{figure}

\begin{figure} [ht]
\centering
\includegraphics[width=0.8\textwidth]{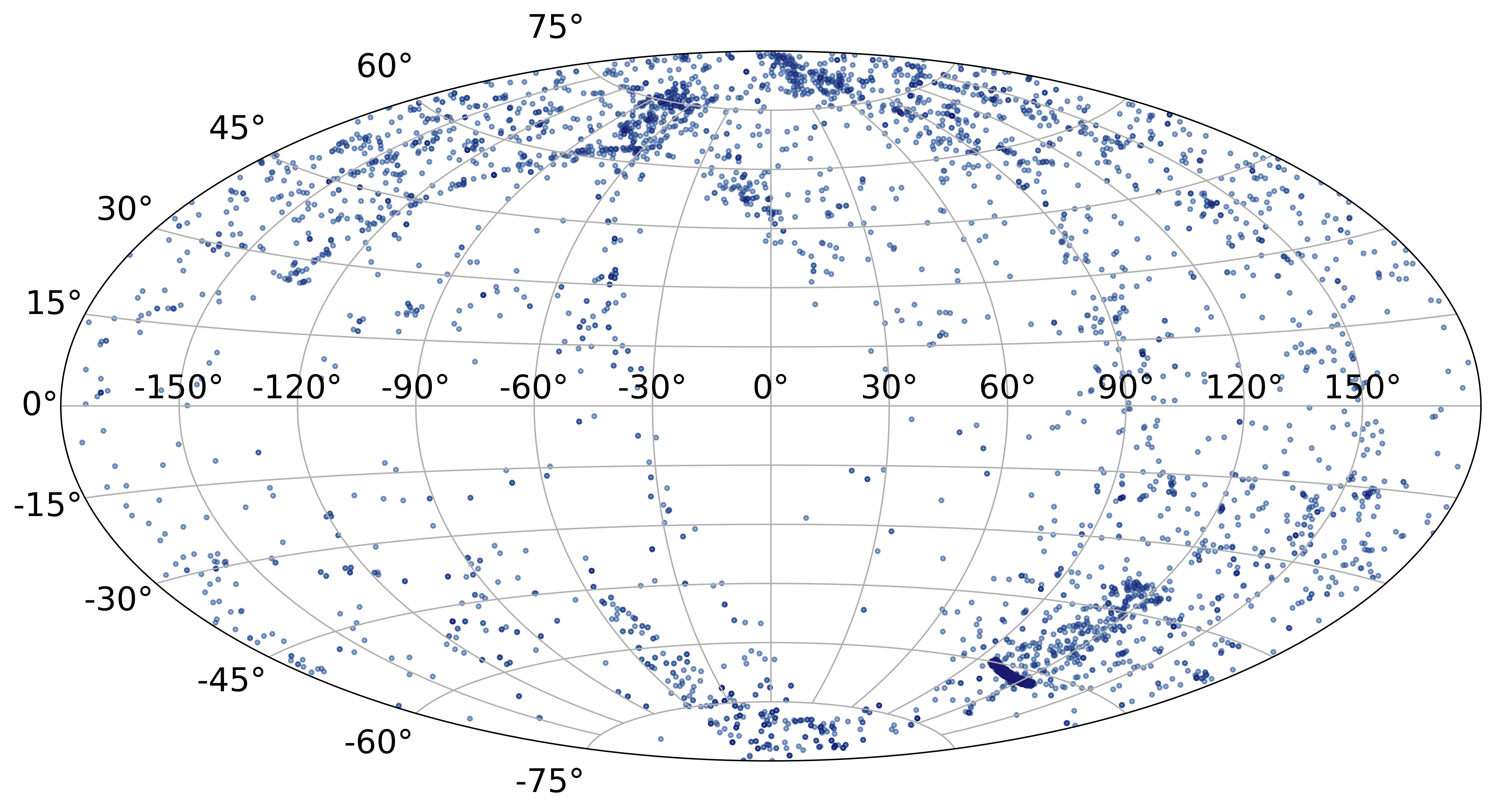}
\caption{Sky locations in Galactic coordinates of the galaxies included in the 43 pointed nearby galaxy surveys included in this review. The  poles are the most densely populated, while the Galactic plane remains largely void of targets due to the additional observational challenges from dust along the line of sight. Data from NGLS, DGS, and the Chandra Nearby Galaxy Survey are not included as individual target coordinates were not include in the survey papers. We refer the reader to the original survey papers for more detailed information.}
\label{fig:2}
\end{figure}

\section{An Overview of Galaxy Surveys} \label{sec: Overview}

\subsection{What is a galaxy survey?} \label{sec: Definition} 
Catalogs of data are not exclusive to recent surveys of galaxies and their evolution, with the earliest records of galaxy surveys being performed with crude observing technology and the unaided human eye. From there came Charles Messier's catalog of nebulae and star clusters \citep{Messier1771}, the New General Catalogue (NGC) in 1888 \citep{dreyer1888NGC}, and the Uppsala General Catalogue (UGC) in 1973 \citep{Nilson1973}. Since then, the definition of a survey has gone through a rapid evolution. Most notably, with the growth of modern observing technology, the number of parameters possible to observe and the resolution of these observations has skyrocketed. Alongside these changes, astronomy has seen the birth of new types of surveys - from large scale legacy surveys consisting of hundreds of galaxies to smaller, very focused surveys of specific populations, such as galaxies in the low surface brightness regime. 

While the impact of new technology on astronomy is clear, the distinction between the types of observing programs is more hazy. What differentiates traditional studies from an atlas or catalog of galaxies, and further still, from a survey? \citet{djorgovski2013} notes that the key scientific distinction between a survey and traditional astronomical observations can be found in the systematic means in which survey data can act as a map of the sky, while also allowing for the characterization of the objects found within an area or within a larger population of objects. This distinction puts an emphasis on the method of choosing the objects or areas to observe, as well as the process of characterizing large quantities of objects in order to extrapolate trends for the greater population. While this definition works well to distinguish traditional observations from larger observing campaigns, it lacks the distinguishing factor between a general catalog of the sky and a sky survey. Most notably, a key difference between catalogs of astronomical objects and sky surveys is the scientific motivation. Catalogs and atlases of the sky are created in order to provide a reference of key characterizations for vast numbers of objects. Sky surveys, however, go beyond solely acting as a reference and aim to probe specific science questions. 

For the purposes of this paper, we have combined the above discussions to form the following definition: a survey is a systematically obtained data set which aims to characterize either all astronomical objects within a portion of the sky, or a defined list of targets, motivated by specific science questions. From this definition, projects which compile previous observations to perform new science, while important contributions to the field, are excluded from our discussion. While large projects which compile observations across multiple wavelength regimes are undoubtedly valuable, an intrinsic property of a survey is the ability for the acquired data to be repurposed for future scientific endeavors. Thus, we felt it necessary to provide a distinction between a survey effort and a project. 

It is common to see the different types of data collections used in tandem -- galaxy surveys will often use a catalog in order to obtain key information on potential targets in order to make a final selection. For example, the H-$\alpha$ Imaging survey of Galaxies in the Local 11~Mpc Volume \citep[11HUGS;][]{kennicutt2008hugs} uses the Third Reference Catalogue of Bright Galaxies to classify the parent sample of galaxies in order to make additional selections. Further, the key components of a galaxy survey are less often the observations (images or spectroscopy) of the targets and, instead, the data products derived from the sample. The observations are then included as a secondary product for potential future reprocessing such as in the Legacy Extragalactic UV Survey \citep[LEGUS;][]{calzetti2015legus} which calculated star formation histories, extinction-corrected ages, and star cluster masses for its sample of targets. 

\subsection{Types of galaxy surveys} \label{sec: Types}
Galaxy surveys can be subdivided into panoramic and pointed galaxy surveys. Panoramic surveys can be classified as surveys which include all objects located within a particular area coverage to a given depth \citep{djorgovski2013}. They offer the most complete pictures of the galaxy population as they observe contiguous areas of sky for prolonged periods of time. By covering significant sky area, and taking observations of all objects within that area, these "all-sky" surveys provide necessary selection bias-free observational samples. These samples of galaxies tend to span a wider set of parameters due to the mix of galaxies located within the field of view. The benefits of random sampling from a certain area of sky include the ability to make predictions on the volume densities of different galaxy types within the local universe. Panoramic  surveys can be further broken down into categories, such as deep or all-sky, based on their coverage range. Panoramic survey data is also often combined to form new projects which follow up on additional science questions, but do not take new observations. The scope of these additional projects is limited by the resolution and accuracy of the original survey observations. As technology allows for large sky coverage surveys to approach the observation quality of traditional targeted surveys, these projects are expected to continue to gain popularity.

On the other end of the spectrum, pointed galaxy surveys observe a sample of defined targets, chosen to fit specific criteria for the science goals of the survey. This results in many pointed galaxy surveys having more homogeneous samples due to the selection criteria imposed. This allows for very specific science questions explored or less common processes, or features, to be studied. Traditionally, having a smaller list of targets meant that the observational cost would be reduced, and higher resolution observations could be taken. This would allow for targets to be studied in greater detail without additional data challenges. Further, some pointed galaxy surveys observed their targets multiple times to characterize their stellar or other variability. In the descriptions below, we note the time domain nature of these surveys but do not otherwise consider them as a separate category. For the remainder of this work, we will use the term 'survey' to refer to a pointed galaxy survey. 

The evolution of galaxy surveys can also be traced through the process of selecting a survey sample. The selection process for galaxy surveys can differ greatly depending on the survey motivation. On one hand, surveys which study large-scale phenomena such as the effects of environment, interactions, or other events on galaxy evolution will select for galaxies based on these extrinsic properties. Conversely, a study targeting a particular feature or property of a population of galaxies (e.g., the Faint Irregular Galaxies GMRT Survey \citep[FIGGS;][]{begun2008figgs}) may select for the particular feature and study the correlation with global parameters. A large limiting factor in early galaxy surveys' sample selection arose from the observational constraints. As technology improved, the available sample of galaxies also increased. Classic reference surveys, such as the Local Volume Legacy Survey \citep[LVL;][]{dale2009lvl}, the GALEX UV Imaging Nearby Galaxy Survey \citep{lee2011galex} or the Herschel Reference Survey \citep[HRS;][]{boselli2010herschel}, required only very broad selection criteria which resulted in complete statistical samples limited only by observing capabilities rather than galaxy parameters. These surveys served as the basis for many other, more focused, galaxy surveys which targeted specific galaxy parameters in order to study individual processes. Examples of this include the Bluedisks Project \citep{wang2013} which studies the disks of unusually HI-rich nearby galaxies, or The Massive Survey which studied the most massive nearby early-type galaxies \citep{ma2014}. Similarly, new access to technology has allowed for observations of select molecular bands, for example The CO Multi-line Imaging of Nearby Galaxies Survey which observed three separate CO bands \citep[COMING;][] {sorai2019co}).  

\section{Survey Sample Selection} \label{sec: Selection}
For the purposes of this review, we have elected to focus on surveys which fit the following selection criteria:

\begin{itemize}
    \item classifies as a pointed galaxy survey (has a defined list of targets selected based on specific selection criteria)
    \item first publication released after 2003
    \item presents new observations or data 
    \item involves a statistically significant sample of galaxies ($>$~10 galaxies)
\end{itemize}

This time period allows for the analysis of the impact that large sky surveys had on the traditional targeted galaxy surveys while maintaining a limited scope. The early 2000s marked the advent of a new era in sky surveys with the launch of several space-based observing campaigns \citep{okamura2020}. The year 2003 saw the first data release from the Sloan Digital Sky Survey \citep[SDSS;][]{SDSS2003}, as well as the launch of the \textit{Spitzer Space Telescope} and the Galaxy Evolution Explorer \citep[GALEX;][]{martin2003}. These new programs were designed in order to be able to observe the entirety of the night sky on shorter timescales, and with higher resolution, than previously seen. These changes opened up the opportunity for a new type of sky survey that had the large, statistical nature of a panoramic survey, while maintaining the precision of traditional targeted survey. We point the reader to \citet{okamura2020} for a more in-depth analysis of recent changes to observational technology and its impact on observational astronomy.

The third criterion stems from the intrinsic property of surveys which is to produce data that has the potential for future use and reprocessing for additional science. We opt to limit our discussion in this paper to surveys which present new observations or data sets in order to focus the scope. We impose a final cut to the sample of surveys which requires a sample size of $>$~10 galaxies, admittedly a somewhat arbitrary choice, which ensures that the results of our sample of surveys are of statistical significance. Table \ref{tab:Tab1} indexes the properties of the surveys included in this project. These surveys span the complete range of wavelengths and select their samples using a wide variety of criteria which we will discuss further in Section~\ref{sec: Discussion}. 

The 43 galaxy surveys included in this report span a wide range of properties. Figure \ref{fig:1} shows the main characteristics of the surveys selected for this project which include: the number of galaxies surveyed, the maximum distance probed, the wavelength domain and the type of the observations (imaging, spectroscopy, or both). UV surveys display the most restricted range of distance coverage with both surveys only extending to approximately 10~Mpc. The X-ray regime has the most targeted surveys with very limited sample sizes. On the other end of the spectrum, radio and optical surveys span the widest range both in the number of galaxies included in samples, and the distance range covered. We also note that there is a clear deficit of both UV and X-ray surveys, while the radio and optical regimes offer dense coverage. Figure \ref{fig:2} presents the distribution on the sky of all the galaxies included in the 43 nearby galaxy surveys of this review. We note that the poles are the most densely populated, while the Galactic plane has sparse coverage due to dust extinction.

\begin{table}[!ht] \label{tab:Tab1}
    \centering
    \begin{tabular}{lllllll}
    \hline
            Name & Year &$N_{\rm gal}$ & $D_{\rm max}$ & $\lambda$ & Type & Parent Catalogue \\ 
                 &      &              & [Mpc]         &           &      & \\ \hline
       \hline
        SINGG & 2005 & 468 & 173.5 & Radio & Img & HIPASS [1] \\ 
        FIGGS & 2008 & 65 & 10 & Radio & Img & Catalog of Neighboring Galaxies [2] \\ 
        THINGS & 2008 & 34 & 15 & Radio & Img & SINGS [3] \\ 
        SHIELD & 2011 & 12 & 20 & Radio & Img & ALFALFA [4] \\ 
        HALOGAS & 2011 & 24 & 24 & Radio & Img & Nearby Galaxies Catalog [5] \\ 
        CHANG-ES & 2012 & 35 & 42 & Radio & Img & Nearby Galaxies Catalog [5] \\ 
        LITTLE THINGS & 2012 & 41 & 10 & Radio & Spect & Hunter and  Elmegreen 2004, 2006 \\ 
        VLA-ANGST & 2012 & 35 & 4 & Radio & Img & ANGST [6], 11HUGS [7] \\ 
        Bluedisks & 2013 & 25 & 130 & Radio & Img & SDSS DR7 [8], GALEX [9]\\ 
        The Massive Survey & 2014 & 116 & 108 & Radio & Both & 2MASS Extended Source Catalog [10] \\ 
        MHONGOOSE & 2016 & 30 & 30 & Radio & Spect & SINGG [11] \\ 
        LeMMINGs & 2018 & 280 & 120 & Radio & Img & Second Reference Catalogue of Bright Galaxies [12], \\ 
                 &      &     &     &       &     & Revised Shapley-Ames Catalogue of Bright Galaxies [13] \\   &      &     &     &       &     & Palomar Nearby Galaxies Survey [14] \\ 
        LVHIS & 2018 & 82 & 10 & Radio & Spect & HIPASS [15] \\ 
        HERACLES & 2009 & 18 & 14 & mm/sub-mm & Spect & THINGS [16] \\ 
        NGLS & 2009 & 155 & 25 & mm/sub-mm & Spect & N/A \\ 
        COMING & 2019 & 147 & 60 & mm/sub-mm & Spect & Nearby Galaxies Catalog [5] \\ 
        PHANGS-ALMA & 2021 & 90 & 20 & mm/sub-mm & Img & N/A \\ 
        SINGS & 2003 & 75 & 30 & IR & Spect & N/A \\ 
        LVL & 2009 & 258 & 11 & IR & Img & ANGST [4], 11HUGS [7] \\ 
        HRS & 2010 & 323 & 25 & IR & Img & N/A \\ 
        S4G & 2010 & 2331 & 40 & IR & Img & HyperLEDA [17] \\ 
        KINGFISH & 2011 & 61 & 30 & IR & Both & SINGS [3] \\ 
        DGS & 2013 & 63 & 191 & IR & Spect & N/A \\ 
        INT & 2020 & 55 & 3 & IR & Img & N/A \\ 
        PHANGS-JWST & 2023 & 19 & 20 & IR & Img & N/A \\ 
        11HUGS & 2008 & 261 & 11 & Optical & Img & Kraan-Korteweg 1986,\\
               &      &     &     &       &     & Nearby Galaxies Catalog [5], \\
               &      &     &     &       &     & Catalog of Neighboring Galaxies [2] \\ 
        ANGST & 2009 & 69 & 4 & Optical & Img & Catalog of Neighboring Galaxies [2] \\ 
        PINGS & 2010 & 17 & 100 & Optical & Spect & N/A \\ 
        CGS & 2011 & 605 & 100 & Optical & Both & Third Reference Catalogue of Bright Galaxies [18],\\
            &      &     &     &       &     & HyperLEDA [17] \\ 
        GHOSTS & 2011 & 14 & 17 & Optical & Img & N/A \\ 
        The ATLAS3D Project & 2011 & 871 & 42 & Multi & Img & 2MASS Extended Source Catalog [10]\\ 
        VENGA & 2013 & 30 & 50 & Optical & Spect & N/A \\ 
        CALIFA & 2014 & 600 & 132 & Optical & Spect & N/A \\ 
        IMSNG & 2019 & 60 & 50 & Optical & Img & GALEX UV Atlas [19] \\ 
        SIGNALS & 2019 & 54 & 10 & Optical & Spect & N/A \\ 
        KMTNet & 2022 & 13 & 50 & Optical & Img & Carnegie Irvine Galaxy Survey [20]\\ 
        PHANGS-HST & 2022 & 38 & 20 & Optical & Img & N/A \\ 
        PHANGS-Halpha & 2022 & 90 & 20 & Optical & Img & N/A \\ 
        PHANGS-MUSE & 2023 & 19 & 20 & Optical & Spect & N/A \\ 
        GALEX UV Img Survey & 2011 & 390 & 11 & UV & Img & 11HUGS [7] \\ 
        LEGUS & 2015 & 50 & 12 & UV & Img & 11HUGS [7] \\ 
        XMM-Newton & 2005 & 27 & 22 & X-ray & Both & Ho et al 1997 \\ 
        CHANDRA & 2005 & 11 & 12.8 & X-ray & Img & N/A \\ \hline
    \end{tabular}
    \caption{A table of the surveys included in this paper and their key properties. [1] \citet{2004HIPASS}, [2] \citet{tully1988nearby}, [3] \citet{kennicutt2003sings}, [4] \citet{giovanelli2005alfalfa}, [5] \citet{karachentsev2004catalog}, [6] \citet{dalcanton2009ANGST}, [7] \citet{kennicutt2008hugs}, [8] \citet{SDSSdr72009}, [9] \citet{martin2003}, [10] \citet{skrutskie20062MASS}, [12] \citet{meurer2006singg}, [13] \citet{1976RC2}, [14] \citet{SandageRSA}, [15] \citet{ho1995}, [16] \citet{walter2008things}, [17] \citet{paturel2003hyperleda}, [18] \citet{1995RC3}, [19] \citet{depaz2007atlas}, [20] \citet{ho2011carnegie}}
\end{table}

\subsection{PHANGS: The Physics at High Angular Resolution Nearby Galaxy Survey} \label{sec:PHANGS}
While there is a general consensus surrounding the crucial role that multi-wavelength observations play in creating a complete understanding of galaxies, their processes and their overall evolution, the approach to obtaining these overlapping observations has varied. One approach is to simply begin with an initial sample of galaxies observed in another survey. Large ``legacy" surveys such as SINGS \citep{kennicutt2003sings}, the ACS Nearby Galaxy Survey Treasury \citep[ANGST;][]{dalcanton2009ANGST} and the Local Volume Survey \citep[LVL;][]{dale2009lvl} are often used as starting points for sample selection due to their large survey samples and general selection criteria. Other surveys make use of large, publicly available archives from specific telescopes to target overlapping observations in specific wavelengths. This is commonly seen in the cases of observationally expensive data sets, for example X-ray data from the \textit{Chandra X-ray Observatory} Public Archive\footnote{\url{ https://cxc.harvard.edu/cda/}} or UV observations from the Galaxy Evolution Explorer Survey (GALEX) \footnote{\url{https://archive.stsci.edu/missions-and-data/galex}}. While these methods offer an efficient use of previous observations, any additional projects are limited by the scope of previous observing campaigns.

To bypass this obstacle, the Physics at High Angular resolution in Nearby Galaxies (PHANGS) Program used simultaneous multi-wavelength observations for a select sample of galaxies \citep{leroy2021phangs}. The PHANGS program is the first program to obtain new observational data in order to study star formation in the nearby spiral galaxy population across multiple wavelengths. Through the combination of numerous large-scale programs such as ALMA, VLT/MUSE, and HST, PHANGS has compiled CO(2-1) imaging, optical spectroscopic mapping, and ultraviolet-optical imaging from each program respectively. PHANGS-JWST is currently ongoing and set to collect mid-infrared imaging for a select subset of galaxies. An X-ray component using the Chandra X-ray Observatory has been proposed \cite{Lehmer2023}. This additional component would target the 19 galaxies in the PHANGS-JWST sample to measure stellar ages and metallicities to high precision.

The parent sample for the program initially included 74 spiral galaxies within a distance range of 4-23 Mpc. These galaxies, originally selected for the PHANGS-ALMA Large Program sample, were determined to be massive ($M_{*} > 10^{9.75} M_\odot$) and star-forming, SFR$/M_* < 10^{-11}$. Further refinements to this criteria were made to ensure that galaxies were not edge-on to the line-of-sight. The parent sample was later expanded to 90 galaxies by incorporating additional nearby galaxies from the ALMA archive for which CO mapping was available. As the criteria used for the PHANGS-ALMA selection process was used as a baseline for all other sub-surveys, the selection criteria are discussed in further detail in Section \ref{sec: PHANGS-ALMA}.  

\section{Radio Surveys} \label{sec: radio}

\subsection{SINGG: The Survey for Ionization in Neutral Gas Galaxies}

The Survey for Ionization in Neutral Gas Galaxies \citep[SINGG;][]{meurer2006singg} was an imaging survey focused on probing the star formation properties of galaxies across the HI mass function. The sample was chosen using a method that was blind to all known optical properties of the galaxies in order to reduce any biases. The project aimed to measure not only the mean star formation quantifiers for the sample, but the distribution of these properties across HI gas mass, Hubble type, surface brightness, and environment parameters. The sample was selected from H I Parkes All Sky Survey \citep[HIPASS;][]{2004HIPASS} using a minimum HI flux to mass ratio of $7.3$~Jy~km~s$^{-1}$ or $1.7\times10^6$~M$_{\odot}$~D$^2$. The final sample included 468 galaxies with a median distance of 18.5~Mpc and a mass range of $8<\log (M/M_{\odot})<10.6$.

\subsection{THINGS: The HI Nearby Galaxy Survey} \label{sec: THINGS}
The HI Nearby Galaxy Survey (THINGS) uses HI emission to survey 34 nearby galaxies \citep{walter2008things}. The high spectral and spatial resolution ($\le 5.3$ km~s$^{-1}$ and $\sim 6\arcsec$ respectively) 21-cm line HI data was obtained using the National Radio Astronomy Observatory (NRAO) Very Large Array (VLA) in order to study the characteristics of the interstellar medium (ISM) and its relationship to galaxy morphology, star formation, and mass distribution. Additionally, as the majority of THINGS targets overlap with galaxies included in SINGS, there is multi-wavelength coverage for most of the sample.
Target galaxies were chosen such that the final sample spanned a large range of physical properties. An exception to this is early type galaxies, which were excluded from the sample due to the lack of neutral atomic gas in their ISM. Additionally, edge-on galaxies were left out of the sample due to the complications of studying the ISM of such systems. Finally, a distance cut was implemented which excludes galaxies in the Local Group and galaxies further than $D = 15$~Mpc to ensure adequate resolution. 

\subsection{SHIELD: The Survey of HI In Extremely Low-Mass Dwarfs}
The Survey of HI In Extremely Low-Mass Dwarfs \citep[SHIELD;][]{cannon2011shield} was focused on studying the neutral gas content of low HI mass galaxies using the Expanded Very Large Array (EVLA) imaging. The 12 galaxies included in the survey have a range of $10^6~M_\odot < M_{HI}< 10^7~M_\odot$ detected from the Arecibo Legacy Fast ALFA Survey \citep[ALFALFA;][]{giovanelli2005alfalfa}. The survey aimed to study the structure, composition, and star formation process in the low mass regime for galaxies outside the Local Group. 
Survey targets were selected from ALFALFA sample. The targets were required to have $M_{HI} \le 1.6\times 10^7 M_\odot$ and a full width at 50\% of peak $< 65$~km~s$^{-1}$ which removed any massive but HI-poor galaxies. The maximum distance of the target sample was 20~Mpc. 

\subsection{HALOGAS: Westerbork Hydrogen Accretion in Local Galaxies Survey}
The Westerbork Hydrogen Accretion in LOcal GAlaxieS (HALOGAS) Survey used deep neutral hydrogen (HI) observations from the Westerbork Synthesis Radio Telescope to study the cold gas accretion in nearby spiral galaxies \citep{heald2011halo}. HALOGAS measured the total HI mass in galaxies, rotation speeds of edge-on systems, and the radial inflow and outflow of gas in galaxies. The HALOGAS Survey provided the deepest publicly available set of HI observations at the time, which allowed for the comparison of gas characteristics and properties of spiral disks in galaxies to provide insight on the global processes such as star formation and gas accretion. 

In order to create a statistically significant sample of target galaxies and reduce selection bias, the criteria for selection were intentionally not based on HI observations. The initial sample of galaxies was adopted from the \textit{Nearby Galaxies Catalog} \citep{tully1988nearby} and then reduced to ensure quality observations. First, only spiral galaxies were chosen to limit the scope in order to achieve completeness. Then, due to the limitations of ground based imaging, a declination limit of $\mathrm{\delta > 25^\circ}$ was imposed. Additionally, to confirm that adequate resolution would be obtained, a maximum distance of 11~Mpc was imposed as well as a limit of $\mathrm{D_{25} >3'}$. From the remaining galaxies, all targets that were found to be of intermediate inclinations or edge-on ($50 ^\circ \le i \le 75 ^\circ$ or i$> 86^\circ$ respectively) were selected.

\subsection{CHANG-ES: Continuum Halos in Nearby Galaxies}
The Continuum Halos in Nearby Galaxies survey aimed to study the halos of edge-on nearby spiral galaxies \citep{irwin2012}. The survey was performed using the Expanded Very Large Array (ELVA) allowing for wide bandwidth observations. The parent sample for the survey was the Nearby Galaxies Catalog \citep{tully1988nearby}. The parent sample was chosen due to its consistency in galaxy parameters, as well as access to local galaxy density measurements. The sample selected for edge-on spirals with inclinations $i>\ang{75}$. In order to be accessible from the ELVA, a declination requirement of $\delta>-\ang{23}$ was imposed. As well, in order to allow for high spatial resolution while maintaining an efficient observing program, galaxy size was limited to blue isophotal diameters within the range of $4^{\prime}<d_{23}<15^{\prime}$. Finally, galaxies were required to have a minimum flux density of $S_{1.4} \geq 23$~mJy at 1.4 GHz to maximize the chances of detection. The final sample of galaxies included both barred- and non-barred spirals across a wide range of environments. The sample also spans a distance range from 2.2 to 42 Mpc.

\subsection{VLA-ANGST: The Very Large Array survey of Advanced Camera for Surveys Nearby Galaxy Survey Treasury galaxies}
The original ANGST survey (see Section~\ref{sec:ANGST}; \citealt{dalcanton2009ANGST}) inspired a follow-up survey which targeted a subsection of original sample. The \textit{Very Large Array survey of Advanced Camera for Surveys Nearby Galaxy Survey Treasury galaxies} \citep[VLA-ANGST;][]{ott2012vla} used high resolution observations of the neutral, atomic hydrogen emission from 35 dwarf galaxies. The program targeted galaxies from the ANGST sample that were not observed in The HI Nearby Galaxy Survey \citep[THINGS;][]{walter2008things} and were observable from the Northern Hemisphere. The maximum distance for the target sample is $\sim 4 \mathrm{Mpc}$. The survey aimed to add to the original ANGST observations while also adopting a similar observational setup to THINGS in order for the two samples to be easily combined. In combination with other surveys that target the low end of the mass spectrum, such as LITTLE THINGS \citep{hunter2012little} and SHIELD \citep{cannon2011shield} (both of which have similar observation strategies), VLA-ANGST provided a broad coverage of local, low-mass galaxies. 

\subsection{The Bluedisks Project}

The Bluedisks project \citep{wang2013} was a focused study of HI rich galaxies in the local universe. The project also compiled a list of control galaxies which were selected based on size in order to be comparable to the original sample. The long-term observing programme used the Westerbork Synthesis Radio Telescope (WSRT) to observe the optically selected sample. The primary goal of the project was to observe the star-forming outer disks of HI rich spiral galaxies in order to probe the differences in morphology, density profiles, and gas kinematics from normal spirals.   

The 25 targets for the Bluedisks project were selected following the optical property-based selection process  by \citet{li2012}. They selected for galaxies expected to be rich in HI gas, as well as a set of galaxies that were comparable in terms of stellar mass, surface density, redshift, and inclination. The starting sample of galaxies came from the Data Release 7 from the Sloan Digital Sky Survey \citep{abazajian2009}. They then selected for galaxies with $10 < \log~M_*<11$, $0.001 <z< 0.03$, and $\delta>\ang{30}$. Additionally, they required the galaxies to have a high signal-to-noise ration in the near ultraviolet based on detection's from the \textit{Galaxy Evolution Explorer (GALEX)} imaging survey \citep{martin2003}. The 1900 successful galaxies then formed the parent sample for the survey from which a sub-sample of 123 galaxies were selected. This secondary selection was based on predicted HI mass fractions in order to gather a sample of HI-rich galaxies. Finally, the survey sample of 25 galaxies was randomly selected from the HI-rich galaxy pool. The final sample spans a distance range of 130~Mpc.

\subsection{The Massive Survey}
The Massive Survey was a volume-limited survey of nearby galaxies \citep{ma2014}. The multi-wavelength survey was designed to study elliptical galaxy formation by studying the spectroscopic and photometric properties of the most massive galaxies within the local universe. Near infrared observations from ground-based telescopes was combined with integral field spectrometry from the Mitchell Spectrograph at McDonald Observatory in order to measure stellar population and kinematics for the survey's sample. 

In order to include the Coma Cluster within the survey area, the maximum distance was set to 108~Mpc, which is larger than previous studies such as $ATLAS^{3D}$ \citep{cappellari2011}. In order to select for the largest galaxies within the distance range, near-infrared \textit{K}-band magnitudes from the Extended Source Catalog \citep{jarrett2000}, which are well correlated with stellar mass measurements, were used. Finally, the sample was limited to elliptical or S0 galaxies in the HyperLeda database \citep{paturel2003hyperleda}. The final sample of 116 galaxies had absolute K magnitude $M_K<-25.3$, declination $\delta>-6^{\circ}$, and distance $D<108~$Mpc.

\subsection{FIGGS: The Faint Irregular Galaxies GMRT Survey}
The  Faint Irregular Galaxies GMRT Survey (FIGGS) uses the Giant Metrewave Radio Telescope (GMRT) to survey a sample of 65 extremely faint nearby dwarf irregular galaxies \citep{begun2008figgs}. The observations include 21 cm neutral hydrogen (HI) observations to study the neutral interstellar medium, supplemented with some incomplete multi-wavelength coverage. These additional observations come from the Hubble Space Telescope (in the \textit{V-} and \textit{I-}bands), and ground based H$\mathrm{\alpha}$ observations from the 6m BTA telescope.

The parent sample used for target selection was the \citet{karachentsev2004catalog} and an initial cut was made to only include galaxies within $\sim 10~$Mpc. The galaxies were also required to have an absolute blue magnitude $M_B>-14.5$ , an HI flux integral $> 1$~Jy~km~s$^{-1}$, and an optical \textit{B}-band major axis $> 1.0^\prime$. The final sample included 65 galaxies within the Local Volume. A follow-up survey (FIGGS2) was performed in 2016 which included 20 additional galaxies \citep{patra2016figss}. Of these 20 galaxies, 15 were observed in the HI 21 cm line using the GMRT.

\subsection{MHONGOOSE: Observing nearby galaxies with MeerKAT}
The MHONGOOSE survey studied the distribution of neutral hydrogen in a representative sample of 30 disk and dwarf galaxies \citep{deblok2016}. The sample spanned a range of HI masses from $10^6M_\odot-10^{11}M_\odot$ and luminosities of $12 \le M_R\le -22$. MHONGOOSE was focused on obtaining both high spectral resolution and column density observations for a representative sample of galaxies in order to probe the effects of cold gas accretion in the local universe, the relation between gas and star formation, as well as the relation between dark and baryonic matter. They were also interested in the distribution of dark matter within galaxies, and the structure, strength and dynamical importance of magnetic fields in galaxies.

MHONGOOSE adopted the 468 targets from the Survey for Ionization in Neutral Gas Galaxies Survey \citep[SINGG;][]{Meurer2006} as the parent sample.  Further criteria were imposed to remove strongly interacting galaxies and galaxies in very dense environments, as well as galaxies with $\delta \le -10^\circ$. Finally, a maximum distance of 30~Mpc was set, which ensured that MeerKAT could achieve the spatial resolution required. The resulting 88 galaxies were divided into 6 bins, from which 5 galaxies per bin were randomly chosen for observation. The chosen galaxies were required to be either edge-on, face-on, or intermediate (defined as having an inclination of 50-60$^{\circ}$. Finally, the sample was selected in order to cover a range in star formation rates and rotation velocities. 

\subsection{LeMMINGs: the Legacy eMERLIN Multi-band Imaging of Nearby Galaxies Survey}
The Legacy eMERLIN Multi-band Imaging of Nearby Galaxies Survey is a radio imaging survey of a subset of galaxies from the Palomar Survey \citep{filippenko1985}. The sample spans a wide range of galaxies including $H_{II}$, Seyfert, LINER, and Transition galaxies. LeMMINGs adopted the galaxy classifications from \citet{ho1997} who selected the galaxies based on optical properties. The primary goal of the survey is to detect the radio component of star-forming or AGN regions on a parsec scale. The survey utilizes a two-tiered observation campaign in which the entire sample is first observed at both 1.5 and 5~GHz for the "shallow" portion. Following these observations, select galaxies were then  imaged a second time, using longer exposure times, for a "deep" survey of scientifically interesting targets. 

The sample observed by the LeMMINGS project inherits a few key characteristics from the Palomar sample. Notably, the parent sample selected galaxies from the Revised Shapley-Ames Catalog of Bright Galaxies \citep{SandageRSA} and the Second Reference Catalogue of Bright Galaxies \citep{deVaucouleurs1976RC2} with $\delta>0^{\circ}$ and $B_T \le 12.5$~mag. The project imposes additional declination limits to the Palomar sample (declination $> 20^{\circ}$) in order to ensure that quality imaging was possible from the eMERLIN array. The final sample spans a distance range out to 108~Mpc and includes over half of the original Palomar sample. 
 
\subsection{LVHIS: The Local Volume HI Survey}
The Local Volume HI Survey (LVHIS) observed deep HI spectral line and 20 cm radio continuum observations for a total of 82 nearby galaxies \citep{koribalski2018lvhis}. These galaxies were selected to be gas-rich and have a distance less than 10~Mpc. The survey aimed to obtain HI gas distributions of galaxies, analyze their structure and gas kinematics, and investigate the influence of a galaxy's environment on the overall HI content and specifically the shape of the outer HI disk. A brief review of the major HI surveys of Local Volume galaxies prior to 2018 can be found in \citet{koribalski2018lvhis}. 

To select a sample for observations, the survey followed \citet{karachentsev2004catalog} in selecting all galaxies with a $V_{\mathrm{LG}} < 550$~km~s$^{-1}$ or independently determined distances of D $<$ 10~Mpc. Then, galaxies with declinations of less than $30^{\circ}$ were isolated to maximize observations. A final additional criterion was applied, which required that the targets must have been detected in the HIPASS survey to ensure that the HI emission was bright enough to allow for a detailed study of the gas distribution and dynamics. The final survey sample consisted of a total of 82 galaxies in the Local Volume. 

\subsection{LITTLE THINGS: The Local Irregulars That Trace Luminosity Extremes, The HI Nearby Galaxy Survey }

The Local Irregulars That Trace Luminosity Extremes, The HI Nearby Galaxy Survey \citep[LITTLE THINGS;][]{hunter2012little} was an in-depth study of the driving factors for star formation in dwarf galaxies. HI observations made with the VLA were combined with UV, optical, and IR data to study the global star formation in the sample of 41 dwarf galaxies. The survey aimed to investigate the factors that regulate star formation in dwarf galaxies, the impact of previous generations of stars on star formation, the relative importance of triggered versus random star formation, and the effects of location within the galaxy on star formation, especially in the outer disks. 

LITTLE THINGS used two previous multi-wavelength studies of dwarf galaxies \citep{hunter2004, hunter2006} as the parent sample. From there, any galaxies with distances $\gtrsim 10$~Mpc were excluded to ensure high resolution observations were possible. They also required the full width at $\mathrm{20\%}$ of the peak of the HI flux-velocity profile $W_{20} > 160$~km~s$^{-1}$. Finally, in following with the original optical survey THINGS (see Section \ref{sec: THINGS}), the sample was selected to cover the entire the full range of integrated properties covered in THINGS. The final sample included 37 dwarf irregular galaxies and 4 blue compact dwarf galaxies within the Local Universe. 

\section{Millimeter and Sub-millimeter Surveys} \label{sec: mm/submm}

\subsection{HERACLES: The HERA CO Line Extragalactic Survey}
The Heterodyne Receiver Array CO Line Extragalactic Survey \citep[HERACLES;][] {leroy2009heracles} was  a CO emission survey that covers 18 nearby galaxies. The HERA multi-pixel receiver on the IRAM 30m telescope was used to map the $CO$ $J = 2\rightarrow1$ transition. The survey includes the full optical disk for each sample target with a resolution of 13\arcsec. The aim was to quantify the relationship between atomic gas, molecular gas, and star formation. The survey included galaxies which were included in the THINGS and SINGS survey samples \citep{walter2008things, kennicutt2003sings}.

HERACLES used the galaxies targeted in THINGS as a starting point for creating the sample for the survey. From there, additional cuts were made to ensure that the targets were far enough north to allow for observation by the 30-m telescope. As well, galaxy targets were required to have an angular diameter of less than $12\arcmin\times12\arcmin$.  The final sample included 18 galaxies of both spiral and irregular morphology, with distances of $\lesssim 14~$Mpc and inclinations of $\le 76^{\circ}$. 

\subsection{NGLS: The JCMT Nearby Galaxy Legacy Survey}
The Nearby Galaxies Legacy Survey \citep[NGLS;][]{wilson2009james} used the James Clerk Maxwell Telescope (JCMT) to create large-area  CO~$J = 3\rightarrow 2$ maps of 155 galaxies with distances between $2-25$~Mpc. When combined with published data, which included CO~$J = 1\rightarrow $ maps, 24 $\mathrm{\mu m}$, and H$\alpha$ images, the survey aimed to study the  properties of the dust and molecular gas in galaxies, and the effect of galaxy morphology and unusual environments on the physical properties of galaxies. 

The galaxies included in the NGLS were initially selected by a HI-flux cut in order to target galaxies with significant interstellar medium while also avoiding given preference to galaxies with high star formation rates. An additional distance cut of 25~Mpc was applied to ensure good spatial resolution. Further, while the sample does include several galaxies from the Virgo cluster, any other galaxies within the Local Group, not within a cluster environment, were excluded. The exception was made for Virgo members to allow for a cluster environment to be studied for completeness. For the non-cluster targets, the HyperLeda database \citep{paturel2003hyperleda} was used to identify galaxies with integrated HI fluxes $< 3.3$~Jy~km~s$^{-1}$. Galaxies with $\delta < -25^{\circ}$ or galactic latitudes between $-25^{\circ}$ and $25^{\circ}$ were also excluded to minimize the effects of Galactic cirrus in previously published data when obtaining multi-wavelength information. For the cluster sample, any Virgo Cluster members with HI flux measurements were included. This resulted in a total of 1150 targets which was ultimately determined to be too large of a sample to observe. 

Further cuts to the galaxy sample included ensuring all SINGS targets which met the above criteria were prioritized (to enhance the overlap in data for each sample). From the total SINGS sample, 47 galaxies fit the criteria and were included in NGLS. An effort was made to include a statistically significant sample in terms of morphological type and environment. A total of four bins were created for morphological type and included E/S0, early-type spirals, late-type spirals, and irregulars. Field galaxies, which had a HI flux measurements of $<6.3$~Jy~km~s$^{-1}$, were assigned a morphological bin and 18 galaxies were randomly selected to be included in the survey sample. For galaxies in a cluster environment, due to the small sample size, the brightest 9 galaxies per morphological bin were included. 

\subsection{COMING: The CO Multi-line Imaging of Nearby Galaxies Survey}
The CO Multi-line Imaging of Nearby Galaxies survey \citep[COMING;][]{sorai2019co} observed a sample of 147 galaxies with $D\lesssim 60~$Mpc using the 45~m telescope of the Nobeyama Radio Observatory. The survey included measurements of line emission in $^{12}CO$, $^{13}CO$ and C$^{18}$ O $J~=~10$ lines. The goal of the survey was to quantitatively advance the understanding of the distribution of molecular gas within a galaxy. 

The COMING team identified 344 objects from the Nearby Galaxies Catalogue \citep{tully1988nearby}, that were bright in the far-infrared (FIR), as FIR flux is well correlated with CO flux. Galaxies with either a $S_{100\mu m} \geq$~10~Jy in the IRAS Catalogue of Point Sources \citep{1988iras} or $S_{100\mu m} \geq 10~$ Jy in the AKARI/FIS All-Sky Survey \citep{Yamamura2010} were selected (targets with dual coverage were required to meet both criteria). Additional criteria excluded any elliptical galaxies due to the challenge of detecting CO emission within the time frame of the survey. Finally, preference was given to objects with previously obtained optical images that had not yet been observed by the 45~m telescope in an attempt to increase the number of CO maps produced within the time limit. The sample also included pairs of galaxies where only one of the objects satisfied the criteria due to the dispersion of the molecular gas within the interacting system. The final sample included 238 galaxies and had a slight bias towards Sb to Sc types due to the restriction of observing conditions.  

\subsection{PHANGS-ALMA}\label{sec: PHANGS-ALMA} 
PHANGS-ALMA is a survey of CO J = $2\rightarrow 1$ (CO(2-1)) emission with a spatial resolution of $\sim$1$\arcsec$ \citep{leroy2021phangs}. The sample was chosen to be representative of nearby galaxies with ``nearby" being defined as $D \lesssim 20$ Mpc. By using CO(2-1) emission, the survey provides a map of the cold-star forming gas of the ISM within nearby galaxies in the local universe. The specific science goals of the PHANGS-ALMA survey include obtaining measurements of the molecular clouds of the sample galaxies, as well as the star formation efficiency and exploring the dependence on galaxy properties.

The galaxies chosen for the sample make up the nearest, massive, star-forming galaxies that are accessible to ALMA. Galaxies with an approximate distance of $D \le 17$ Mpc were targeted as this ensured that the galaxies themselves were close enough that $ 1\arcsec \le 100$ pc. Further, only galaxies that did not have extreme inclinations ($i < \ang{75}$), that fit the classification as massive (defined as having a stellar mass $log_{10} (M_*/M_{\odot}) \gtrsim 9.75$), and are actively star forming (a specific star formation rate (sSFR) $> 10^{-11}$~yr$^{-1}$) were considered. Finally, the targets were required to be visible to ALMA and thus only galaxies with a declination between $\delta = -75^{\circ}$ and $\delta = +25^{\circ}$ were targeted. 

\section{Infrared Surveys} \label{sec: IR}

\subsection{The Spitzer Nearby Galaxy Survey} \label{sec: SINGS}
The Spitzer Nearby Galaxy Survey (SINGS) was an infrared imaging and spectroscopic survey with a primary goal of characterizing the infrared emission of galaxies and their primary infrared-emitting components \citep{kennicutt2003sings}. The survey sample spanned a wide range of galaxy properties and environments, and included 75 galaxies in the local universe. The survey included imaging from both IRAC ($3.6\mathrm{\mu m}$ and $ 4.6\mathrm{\mu m}$) and MIPS. SINGS used a 2-step approach in their data collection strategy. First, the survey sample was imaged with IRAC and MIPS which allowed for the overall infrared and star-forming properties to be characterized. Following the imaging observations, additional spectroscopic observations were also taken for the complete sample.
SINGS consists on a diverse set of 'normal' galaxies within the local volume (30~Mpc). Further, as most of IRAC and MIPS images have a field of view of 5' due to the observing constraints from IRAC and MIPS, galaxies with sizes in the 5$\arcsec$-15$\arcsec$ range were selected to ensure efficient use of the observing time allotted. Additionally, galaxies near the Galactic plane were excluded to minimize background interference. To create a representative sample of galaxies, roughly equal numbers of each galaxy type from the Third Reference Catalogue of Bright Galaxies 
(RC3) were chosen. This final sample covers a factor of $10^5$ in infrared luminosity and is representative of a population of \textit{normal} local galaxies. 

\subsection{LVL: The Local Volume Legacy Survey}
The Spitzer Local Volume Survey \citep[LVL;][]{dale2009lvl} provided infrared imaging for a large sample of galaxies within the Local Volume. The survey was intended to have a large overlap with previously published data and surveys to allow for multi-wavelength observations. LVL obtained near-, mid-, and far-infrared flux measurements (IRAC (3.6, 4.5, 5.8, and 8.0 $\mu$m) and MIPS (24, 70, and 160 $\mu$m)) for a total of 258 galaxies. The chosen sample had previous observational data in H$\alpha$, UV (from GALEX), and optical (from HST). With the combined data set, LVL aimed to provide insight into the relationship between infrared emission, dust content, and global galaxy properties; as well as the physical mechanisms related to dust heating and the factors which influence polycyclic aromatic hydrocarbon (PAH) emission. 

The LVL sample is comprised of two parts: the inner sample which goes out to a radius of $3.5$~Mpc and the outer sample which extends to $\sim 11$~Mpc. The inner sample was based on the sample of galaxies targeted in ANGST (see Section~\ref{sec:ANGST}) and included all known galaxies out to the distance cut, excluding the Local Group and any galaxies which lie within the Galactic plane ($|b| > 20^{\circ}$). The inner sample was covered by GALEX UV imaging and deep HST optical imaging, allowing for multi-wavelength analysis. The outer sample followed the $H\alpha$ survey and the GALEX UV imaging (collectively known as 11HUGS, \citealt{kennicutt2008halpha}) with slightly loosened criteria, and included galaxies out to a distance of 11~Mpc, with $|b| > 20^{\circ}$ and $m_B < 15.5$~mag. 

\subsection{The Herschel Reference Survey}
The \emph{Herschel} Reference Survey \citep{boselli2010herschel} was a benchmark study of the dust in the nearby universe. It contained a statistically complete sample of 323 galaxies that spanned a wide range of morphological types and environments. The survey aimed to accomplish a total of nine science goals focused on investigating the properties of dust content and the effects on the global parameters of the galaxy, and characterizing the origin and evolution of dust in galaxies. 

The \emph{Herschel} Reference Survey's sample contains over 300 galaxies within a distance range of $\sim15-20$~Mpc. The lower limit imposed on the distance excludes nearby sources that are overly extended in order to maximize observation efficiency, while still including the Virgo cluster to study the effects of a dense environment. Further limits on K-band emission were adopted to minimize selection bias in galaxy types by applying different cutoff criterion to early versus late type galaxies. Finally, to minimize the contamination due to the galactic cirrus, the survey selected for high galactic latitude ($b > 55^{\circ}$) and low galactic extinction ($A_B < 0.2$). Of the final sample included in the survey, 65 of the 323 galaxies are early-type, while the remaining 258 are late-type. 

\subsection{S4G: The Spitzer Survey of Stellar Structure in Galaxies}
The Spitzer Survey of Stellar Structure in Galaxies (S$^4$G) used the Infrared Array Camera (IRAC) to explore the stellar structure of 2331 galaxies in the local universe \citep{sheth2010spitzer}. The sample was observed at both 3.6~$\mu$m and 4.5~$\mu$m  and mapped to an angular diameter of $\ge 1.4 \times D_{25}$. Further, the sample of targeted objects contained all Hubble types including dwarfs, spirals, and ellipticals. Measurements of the diameter, position angle, axial ratio, and inclination were made for each galaxy in the sample, as well as the total magnitude, ellipticity, surface brightness and color.

S$^4$G sample targets were chosen to have radial velocities of $< 3000$~km~s$^{-1}$ which corresponds to a distance of $<40$ Mpc for a Hubble constant of $75$~km~s$^{-1}$~Mpc$^{-1}$. This criterion does limit the sample to galaxies with radio-derived radial velocities and misses some potential targets with only optically derived radial velocities. The program did compare the potential sample defined by optical radio velocities and noted that the galaxies excluded from the radio-derived sample were systematically small, relatively faint and gas-poor. Additional criteria include: a total corrected blue magnitude $< 15.5$, a blue light isophotal angular distance $> 1.0\arcmin$, and Galactic latitude $|b| >30^{\circ}$. The volume cut of 40~Mpc was chosen such that it would be large enough to include a statistically significant sample of galaxies which included galaxies of all types and environments. Within the sample, 188 targets are classified as early-type galaxies, 206 are dwarf galaxies, and 465 are edge-on systems (with $i > \ang{75}$). Due to the large number of targets, there is significant overlap with other surveys. In particular, 125 of the galaxies overlap with the Local Volume Legacy (LVL) survey and another 56 are in common with SINGS. 

The sample used in \citet{sheth2010spitzer} was selected using velocity cuts derived from radio measurements alone. The result of this was a systematic exclusion of gas-poor, primarily early-type galaxies. In order to create a more representative sample which included more early-type galaxies, the S$^4$G Team carried out an additional survey which followed identical methods to the original survey on a sample of 465 early type galaxies \citep{sheth2013}. \citet{Watkins2022} continued the survey by processing the additional data using the original pipeline to conduct the missing photometric analysis. This included deriving surface photometry for each new early-type galaxies, stellar masses, and effective radii. The new sample of galaxies was about two thirds S0 type galaxies, while the rest were classified as elliptical galaxies, and was selected from the HyperLEDA database \citep{paturel2003hyperleda}. The added sample also included 7 smaller satellites of the Milky Way. The other parameters of the sample were similar to those of the original S$^4$G survey with a maximum distance of 40~Mpc and a galactic latitude $|b|\geq 30\deg$, however the final sample had a slightly higher median stellar mass of $log(M_*/M_{\odot})$ = 10.15. 

\subsection{KINGFISH: Key Insights on Nearby Galaxies: a far-infrared survey with \textit{Herschel}}

The Key Insights on Nearby Galaxies: a Far-Infrared Survey with \textit{Herschel} \citep[KINGFISH;][]{kennicutt2011king} included both imaging and spectroscopic observations of 61 nearby galaxies ($\le 30~$Mpc). The survey aimed to fill the gap in star formation observations that is caused due to dust effects, probe the relation between the processes of star formation and dust heating, investigate the dust components for galaxies, in particular the cold gas component, and study the temperature variances in the ISM. KINGFISH was specifically designed to be combined with SINGS (see Section \ref{sec: SINGS} to create a complete galaxy sample for observations. The KINGFISH observations include both  broad-band imaging at 70, 100, 160, 250, 350, and 500 $\mu$m and spectral line mapping in the far-infrared emission lines ([O I] 63~$\mu$m, [O III] 88~$\mu$m, [N II] 122,205~$\mu$m, and [C II] 158~$\mu$m) and were designed to be comparable with those from SINGS. From the 75 galaxies that were targets in SINGS, ten of the galaxies had already been allocated observation time in other \textit{Herschel} Guaranteed Time Programs. From the remaining galaxies, 8 others were excluded due to having very similar properties to others in the sample, in order to boost observational efficiency. In order to extend the range of properties covered in the sample, an additional 4 galaxies from other \textit{Spitzer} surveys were added. The final sample contained 61 galaxies with distances of $\le 30$~Mpc and covered a wide range of properties. 

\subsection{The Dwarf Galaxy Survey}
The Dwarf Galaxy Survey \citep[DGS;][] {madden2013dwarf} focused on nearby star-forming dwarf galaxies in order to investigate the global properties of low metallicity galaxies, the impact of star formation process on the interstellar medium, and the roles and relationships between gas and dust in galaxies. The galaxies were observed using the \textit{Herschel Space Observatory} with imaging at 70, 100, 160, 250, 350, and 500~$\mu$m and spectral line mapping in the far-infrared emission lines ([NIII] 57~$\mu$m, [O I] 63~$\mu$m, [O III] 88~$\mu$m, [N II] 122,205~$\mu$m, [OI] 145~$\mu$m and [C II] 158~$\mu$m), and high-resolution spectroscopy (194-–671~$\mu$m) for a few targets. The targeted galaxies have previous observations across a range of wavelengths which can be used in future multi-wavelength projects. The maximum distance of the same is $\sim 200$~Mpc. The DGS aimed to compile a statistically significant sample of dwarf galaxies across a large metallicity range; the final sample included 50 galaxies. 

\subsection{INT: The Isaac Newton Telescope Local Group Dwarf Galaxy Survey}
The Local Group Dwarf Galaxy Survey used the 2.5m Isaac Newton Telescope to image 55 dwarf galaxies within the local group, as well as 4 globular clusters \citep[INT;][]{saremi2020}. The survey aimed to identify asymptotic giant branch (AGB) stars, and red super giants at the ends of their evolution in order to study the star formation histories of local dwarf galaxies. They also hope to identify long period variable (LPV) stars. The survey used the Wide Field Camera and the Sloan \textit{i} and Harris \textit{V} filters chosen to maximize the contrast between LPVs and other stars. 

The sample of galaxies were required to be visible in the Northern Hemisphere. Priority was given to Andromeda satellites in order to create a uniform sample in terms of distances, completeness, classification accuracy, and observation quality. The remaining galaxies in the sample were chosen based on estimations of AGB star numbers. The final sample of 55 dwarf galaxies spanned the range of morphologies and included 43 dSphs, six dIrr, and six dTrans in addition to the 4 globular clusters.

\subsection{PHANGS-JWST}
The PHANGS program expanded its wavelength coverage to the IR in 2023 \citep{lee2023phangs}. The PHANGS-JWST program aims to target 19 of the parent sample galaxies that have overlapping data from PHANGS-ALMA, MUSE, and HST in order to obtain imaging data ranging from 2-21$\mu$m. Analysis of this data is being preformed with specific focus on the following: characterizing the young star populations and the ISM bubble and shell features, measuring the duration of dust embedded star formation and the efficiency of star formation, determining the effects of local interstellar conditions on the properties, evolution, and processing of small dust grains (ex. polycyclic aromatic hydrocarbons), developing new dust-based tracers of the neutral gas, and establishing a robust mid-IR diagnostic of star formation activity in order to account for young stars. The PHANGS program has been proposed to expand the project to include 554 more galaxies in a PHANGS JWST Cycle 2 Treasury Program \citep{williams2024}.

\section{Optical Surveys} \label{sec: optical}

Optical surveys offer the most variation in terms of the motivating science goals and observing projects. Optical surveys can be further split into categories based on the observing methods utilized. These categories include imaging (static or synoptic monitoring) and spectroscopy (integrated or spatially resolved). While synoptic spectroscopic monitoring surveys of galaxies could in principle be performed, our search of the literature did not uncover any.

\subsection{11HUGS: An H$\alpha$ Imaging Survey of Galaxies in the Local 11Mpc Volume} 
The H$\alpha$ Imaging survey of Galaxies in the Local 11Mpc Volume \citep[11HUGS;][] {kennicutt2008hugs} observed a sample of 261 galaxies to create a catalog of integrated H$\alpha$ fluxes, luminosities, and characteristics of the galaxy a statistical sample. The survey aimed to understand the process of star formation by characterizing the full range of star formation properties for galaxies in the Local Volume and constraining the behaviour of star formation in dwarf galaxies. The galaxy sample used in 11HUGS was used as the basis for other legacy surveys using GALEX \citep{lee2011galex} and \textit{Spitzer} Local Volume Survey \citep{dale2009lvl}. 

By defining a maximum distance, the 11HUGS team was able to identify a complete sample of galaxies while also maintaining an efficient observing program. Poor measurements were minimized by eliminating galaxies near the galactic plane ($b <20^{\circ} $), and requiring a minimum brightness $B\le 15$. Further, primary sample was limited to spiral and irregular galaxies (as classified in the third edition of the \textit{Reference Catalogue of Bright Galaxies}, \citep{1995RC3}) while elliptical, dwarf spheroidal, or gas-poor S0 galaxies were rejected due to their limited detectable HII regions. 

\subsection{ANGST: The ACS Nearby Galaxy Survey Treasury} \label{sec:ANGST}
The ACS (Advanced Camera for Surveys) Nearby Galaxy Survey Treasury \citep[ANGST;][]{dalcanton2009ANGST}, began in 2006 and aimed to compile a uniform multi-color photometry of resolved stars for a select sample of nearby galaxies. Images in the F475W, F606W and F814W bands were taken with the Advanced Camera for Surveys (ACS) and later the Wide Field Planetary Camera (WFPC, following the ACS failure) on the \textit{Hubble} Space Telescope (HST). The survey sample included 69 galaxies spanning a diverse range of environments, from isolated regions to large groups and filaments. The properties of the galaxies within the sample also covered a complete range, with luminosity and star formation rate measurements ranging across a factor $\sim10^4$. The vast range covered by the survey, as well as the care taken to take an unbiased sampling of the local universe, allows for the ability to make comparisons between galaxies within the sample, as well as cosmological simulation data. Sample galaxies were chosen to balance creating a complete sample in order to achieve scientific goals while maintaining an efficient observational program. 

Beginning with the \citet{karachentsev2004catalog} Catalog of Neighbouring Galaxies as an initial sample, the catalog was then reduced to only include galaxies beyond the zero velocity surface of the Local Group. This choice was made to maximize efficiency of ground-based observations and to include the large number of existing HST observations.  An initial maximum distance of 3.5~Mpc was adopted, however, as the Local Volume at this range contains mostly field galaxies, the distance was expanded to $\sim$ 4~Mpc to include the massive M81 group (D $\approx 3.6$~Mpc) and NGC 253 clump (D $\approx$ 3.9~Mpc) for completeness in environment density and luminosity. This volume cut resulted in a total of 69 galaxies that cover a large luminosity range in both the B- and K- bands, and contains a wide variety of morphological types residing in diverse environments. The morphological type break down of the sample is as follows: $17\%$ of the sample are characterized as spirals, $ 25\%$ are classified as dwarf ellipticals and $58\%$ are dwarf irregulars.  Further, galaxies selected were required to have $|b| >20^{\circ}$ to avoid incompleteness of the sample at low galactic latitudes. 

\subsection{PINGS: The PPAK IFS Nearby Galaxy Survey}
The PPAK IFS Nearby Galaxies Survey (PINGS) was a two-dimensional spectroscopic survey which includes 17 nearby spiral galaxies \citep{rosales2010pings}. The survey targeted the entire surface of the galaxies in the optical wavelengths and covered a total area of $\sim 80$ $\mathrm{arcmin^2}$. The spectroscopic data was complemented with previous broad-band and narrow-band imaging. The Potsdam Multi Aperture Spectrograph fibre PAcK (PPAK) at the Centro Astronomico Hispano Aleman has a field of view of $74 \times 65$ $\mathrm{arcsec^2}$ and provided 2D spectroscopic mosaics for the entire galaxy sample, covering wavelengths from 370--710~nm with a spectral resolution of $\sim 1$~nm full width at half-maximum. The main goal of the survey was to observe the variation in the line emission and stellar continuum of the disks of nearby galaxies.

In order to achieve statistically representative results, the sample of galaxies was carefully chosen to cover a range of different galaxy types. Some limitations due to the FOV of the PPAK unit, and the required observing time, caused some bias in the size and natures of the galaxies included in the sample. Preference was given to face-on galaxies with high surface brightness and active star formation. Further, the location of the galaxy target was required to be observable by the Calar Alto observatory. The final sample of 17 galaxies spanned a variety of morphological types, environments, and galactic properties. Additionally, effort was taken to ensure some overlap with the SINGS survey \citep{kennicutt2003sings} which allows for complementary UV, infrared, HI and radio data. 

\subsection{CGS: The Carnegie-Irvine Galaxy Survey}
The Carnegie-Irvine Galaxy Survey \citep[CGS;][]{ho2011carnegie} used both photometry and spectroscopy to observe a statistically complete sample of nearby bright galaxies in the Southern Hemisphere. The survey used the du Pont 2.5 m telescope to target 'bright' galaxies with $B_T < 12.9$~mag. No preference was given to morphology, size, or environment while selecting targets. The final sample of galaxies consisted of 605 galaxies which covered a distance range of almost $100$~Mpc and the full range of Hubble types that exist in the Local Volume. Finally, while a few non-isolated systems were included, the majority of the galaxies surveyed fell under the classification of 'undisturbed'. 

\subsection{GHOSTS: Galaxy Halos, Outer disks, Substructure, Thick disks, and Star clusters Survey}
The Galaxy Halos, Outer disks, Substructure, Thick disks, and Star clusters Survey \citep[GHOSTS;][]{radburn2011ghosts} was a survey of the outer regions of disk galaxies and specifically targeted their the resolved stellar populations. The sample contains 14 disk galaxies whose outer disks and halos were imaged with  HST/ACS and WFPC2. The survey used two filters to image the sample: F814W, which targets RGB stars which were expected to be abundant within the halos of the sample targets, and F606W, which is used for color discrimination.

The initial sample came from the list of undisturbed disk galaxies from \citet{karachentsev2004catalog} with a distance range of 1--5~Mpc and rotation velocities $> 90$~km~s$^{-1}$. The small distance range allowed for a signal-to-noise ratio $>10$ to be achieved, even with short exposure times of the HST SNAPshot observations. Additional edge-on galaxies ($i < \ang{87}$) were targeted which allowed for the galaxy halos to be studied, undisturbed by the disk. Further cuts were made to the sample to require galaxies to have rotation velocities $>80$~km~s$^{-1}$ and distances $\lesssim 14$~Mpc. This resulted in a combined sample of 14 galaxies.

\subsection{The ATLAS-3D Project}
The ATLAS-3D Project \citep{cappellari2011} was a volume-limited survey of 260 elliptical and lenticular galaxies. It combined existing radio, millimetre, near-infrared and optical imaging data with new integral field spectroscopy from the SAURON instrument on the William Herschel Telescope (WHT). The SAURON field of view is $33\times 41$~arcsec, sampled with lenslets of area 0.94 arcsec$^2$. Spectral resolution for the ATLAS-3D observations was 0.42~nm FWHM, covering a wavelength range 480–538~nm. The targets were selected from a parent sample of 871 galaxies within 42~Mpc that are observable from WHT, have Galactic latitude $|b|>15\deg$ , and have total $K$-band magnitudes from the 2MASS Extended Source Catalog \citep{jarrett2000} of $M_K<-21.5$ (corresponding to $M_*>6\times 10^{9}$~M$_{\sun}$). The authors analyzed possible selection biases and concluded that the parent sample is essentially complete and statistically representative of the nearby galaxy population. Science goals of ATLAS-3D included studies of star formation history, kinematics, and stellar populations in elliptical and lenticular galaxies, with goals of linking the properties of present-day galaxies to those of higher-redshift galaxies.

\subsection{VENGA: The VIRUS-P Exploration of Nearby Galaxies}
The VIRUS-P Exploration of Nearby Galaxies (VENGA) was an integral-field spectroscopic survey of nearby spiral galaxies \citep{blanc2013venga}. VENGA used the VIRUS-P spectrograph on the 2.7 m Harlan J. Smith telescope at the McDonald Observatory. Wavelength coverage was 360--680~nm with spectral resolution of 0.5~nm FWHM, with data cube spatial resolution of 5\farcs6 FWHM.
\citet{blanc2013venga} compared VENGA to other integral-field spectroscopic surveys such as PINGS and CALIFA, and noted that VENGA has a similar wavelength coverage, factor of two higher spatial resolution, and is ``a factor of a few deeper.’’
Quantitative selection criteria are not explicitly described by \citet{blanc2013venga}. The sample of 30 northern spiral galaxies was chosen to span a range in morphology, inclination, and right ascension. It includes galaxies with distances $4<D<50$~Mpc, star formation rates from 0.2 to 39 M$_{\sun}$~yr$^{-1}$, and stellar masses $8\times 10^8 < M_*<3\times10^{11}$~M$_{\sun}$. Most of the target galaxies are Milky Way size or larger:  28 of 30 have $M_*>10^{10}$~M$_{\sun}$.  Science goals of the survey included studies of the interstellar medium, star formation, stellar populations, galaxy kinematics and mass assembly, and chemical evolution.

\subsection{CALIFA: Calar Alto Legacy Integral Field Area Survey}
The Calar Alto Legacy Integral Field Area Survey (CALIFA) provided IFS observations for 600 galaxies in the nearby universe \citep{walcher2014califa}. Wavelength coverage was 370--750~nm with spectral resolution of 0.6~nm full width at half maximum; a `remastered' dataset has recently been released \citep{sanchez2023}. By taking an initial sample of galaxies from the Sloan Digital Sky Survey (SDSS) and basing the selection of galaxies purely on visibility limits, the survey attempted to maintain the statistical properties of the SDSS sample while filling in a gap in IFS observations over a luminosity range of $-19~\le M_R~\le -23.1$ and a stellar mass between $10^{9.7}~\le M_*/M_\odot~\le 10^{11.4}$. 

To maintain a statistical sample while completing an efficient observational study, the CALIFA sample was randomly selected from a larger parent sample. This sample was chosen using only criteria that would ensure quality observations. Galaxies were required to have a minimum size (an isophote semi-major axis at 25 mag~arcsec$^{-2}$ in the R-band between 45-79.4\arcsec), be offset from the galactic plane, and within a distance range of $\mathrm{\sim} 20-130$~Mpc. To further minimize any observational challenges, the hour angle of the targets was restricted to $\mathrm{-2h~\le HA~\le~2h}$ and the declination to $\delta > \ang{7}$ in order to reduce the effects of atmospheric refraction. Final visual adjustments were made to the resulting 942 galaxies to exclude any galaxies with poor SDSS imaging or galaxies lacking redshift measurements.

\subsection{IMSNG: The Intensive Monitoring Survey of Nearby Galaxies}
The Intensive Monitoring Survey of Nearby Galaxies \citep[IMSNG;][]{im2019} was an imaging survey focused on monitoring supernovae in nearby galaxies. By observing the sample of 60 galaxies with a high observing cadence on the order of hours, the survey aimed to investigate the explosion mechanism in supernovae. The sample of galaxies was chosen to maximize the probability of observing supernovae. This required galaxies with high star formation rates to be targeted. In order for these galaxies to be identified, a minimum limit was placed for the near UV luminosity which is well correlated with star formation rates \citep{kennicutt1998}. For this, galaxies with $M_{NUV}<18.4 $ mag were selected from the GALEX UV Atlas \citep{depaz2007atlas, bai2015}. A maximum distance of 50~Mpc was imposed, as well as limit of b$~< 20^{\circ}$ in order to limit the effects of Galactic extinction. The final sample was composed of 60 galaxies including 22 AGNs, 11 Seyfert galaxies, 10 low-ionization nuclear emission line regions (LINERs). 

\subsection{SIGNALS: The Star-formation, Ionized Gas, and Nebular Abundances Legacy Survey}
The Star-formation, Ionized Gas, and Nebular Abundances Legacy Survey (SIGNALS) is an observing program aimed at probing massive star formation and HII regions in 54 nearby galaxies \citep{rousseau2019signals}. The program uses the Fourier transform spectrograph SITELLE on the Canada-France-Hawaii Telescope (CFHT) to observe in three wavelength ranges (SN1: 363--386 nm, SN2: 482--513 nm, and SN3: 647--685 nm) with spectral resolutions of $\lambda/\Delta \lambda$ of 1000, 1000, and 5000, respectively, and a mean spatial resolution of $\sim 20$~pc. The specific science goals of SIGNALS include quantifying the impact of environment on star formation and measuring variations in the resolved star formation rate with respect to various indicators.

The sample selection for SIGNALS is driven by the goal of observing numerous HII regions in a variety of different environments. They identified star forming regions using existing H$\alpha$ and UV-GALEX images for nearby galaxies observable by CFHT (-\ang{22} $< \delta < +\ang{62}$) to select sample targets. Further, a maximum distance range was set to be 10 Mpc to optimize the spatial resolution of the observations. Finally, targets are required to have a limited amount of dust on the line of sight and limited crowding of the HII regions to ensure clear observations. Using these criteria, a sample of 54 objects was identified and targeted. The sample covers an absolute magnitude range from $\sim -21.3$ to $\sim -13.5$ and a wide range of galactic environments, including 1/3 of the targets being isolated objects and the other 2/3 in groups with no strongly interacting systems. The targets chosen include 6 objects overlapping with PHANGS-MUSE (and thus PHANGS-ALMA, H$\alpha$, and HST as well). 

\subsection{KMTNet: The Korea Microlensing Telescope Network Nearby Galaxy Survey}
The Korea Microlensing Telescope Network Nearby Galaxy Survey \citep[KMTNet;][]{byun2022kmtnet} includes deep and wide-field images of 13 nearby galaxies in the Southern Hemisphere. The survey targets the outer regions of spiral galaxies with low-surface-brightness features. The field of view for the survey is $\sim$ 12 $\mathrm{deg^2}$ and images were taken in both the optical broad bands (\emph{B, R, I}) and an H$\alpha$ narrowband. The survey targets galaxies with disks containing extended UV (XUV) emission to explore active star formation in the outer disk area. 

The initial sample of targets was taken from the Carnegie-Irvine Galaxy Survey \citep[CGS;][]{ho2011carnegie} which included 605 galaxies in the Southern Hemisphere. Galaxies with XUV emission in the outer regions of the disk, identified by \citet{thilker2007search} using both far- and near- UV images from the GALEX satellite, were then matched with the CGS sample and targeted. All targets are within $\sim$ 50 Mpc which maximizes the wide field-of-view of the KMTNet. 

\subsection{PHANGS-HST}
The PHANGS-HST imaging program included 38 galaxies from the PHANGS-ALMA parent sample \citep{lee2022phangs}. These galaxies were determined to be the best suited for the joint program analysis, specifically for the resolved population of young stars and giant molecular clouds. The motivation behind the survey was to use five-filter broad-band imaging to help locate and classify stellar clusters within the local universe. Further criteria were imposed on the parent sample which required galaxies to have inclinations $\le \ang{70}$, avoid the Galactic plane ($|b| < 15\arcdeg$) to minimize the effects of foreground stars, and to be star forming (defined as having SFR $\gtrsim 0.3~{\mathrm M}_{\odot}$~yr$^{-1}$). This resulted in a final sample of 38 galaxies for the HST observing program. These galaxies include spiral galaxies that span morphological types Sa to Sd, SFR from $\sim$ 0.2-17~${\mathrm M}_{\odot}$~yr$^{-1}$. The imaging taken for the 38 galaxies aimed to cover the regions previously mapped by ALMA in five different filter: NUV, U, B, V, and I (or F275W, F336W, F438W, F555W, and F814W). The galaxy halos were simultaneously observed using the Advanced Camera for Surveys Wide  Field Channel (ACS/WFC). The PHANGS-HST sample includes all 19 of the galaxies mapped by the PHANGS-MUSE program.

\subsection{PHANGS-MUSE}
PHANGS-MUSE is a program aimed at mapping the properties of ionized gas and stellar populations in the discs of 19 massive galaxies selected from the PHANGS-ALMA parent sample \citep{emsellem2022phangs, groves2023phangs}. Observations with the VLT/MUSE integral field spectrograph cover a wavelength range 475--935~nm, with spectral resolution 80 km~s$^{-1}$ (red) to 35 km~s$^{-1}$ (blue), over multiple pointings per galaxy. The PHANGS-MUSE galaxies cover a wide range of stellar masses, but are somewhat biased toward galaxies on the main-sequence. This sample of galaxies was initially targeted in an effort to align with the PHANGS-ALMA pilot survey. A mass cut was applied to the parent sample in an effort to target only massive star-forming galaxies (defined as having $\log(M_*/M_{\odot})\gtrsim 9.75$ and $\mathrm{sSFR} > 10^{-11}~yr^{-1}$). The additional criteria implemented on the parent sample ensures that the molecular clouds and regions of star formation within the targets of the program can be isolated enough to map. 

\subsection{PHANGS-H\(\alpha\)}
In order to compliment the PHANGS-MUSE program, narrowband (1$\arcsec$) H\(\alpha\) imaging was obtained by the PHANGS-H\(\alpha\) program for the entire parent sample \citep{lee2022phangs}. This allowed for SFR maps to be obtained, as well as catalogs of ionized nebulae. The program used the Wide Field Imager (WFI) on the ESO/MPG 2.2m telescope in combination with the DirectCCD on the du Pont 2.5m telescope.

\section{Ultraviolet Surveys} \label{sec: UV}
\subsection{A GALEX UV Imaging Survey of Galaxies in the Local Volume}
The GALEX UV Imaging Survey of Galaxies in the Local Volume Survey targeted 390 galaxies within 11~Mpc \citep{lee2011galex}. The survey sample is dominated by dwarf galaxies with luminosities and star formation rates lower than that of the Local Magellanic Cloud (LMC). Using the GALEX satellite, the survey compiled both FUV and NUV observations (150~nm and 220~nm respectively), and computed two separate measures of the global star formation efficiency (SFE) for the sample's targets. The goals of the survey included providing a statistically complete UV data set for a deep Local Volume galaxy sample in order to participate in the larger multi-wavelength initiative to study the mechanisms that drive the cycle of star formation on a global scale in galaxies, with a particular focus on dwarf low surface brightness galaxies. 

The GALEX UV Imaging Survey used a parent sample taken from 11HUGS \citep{kennicutt2008hugs} and thus overlaps with both the 2007 GALEX UV Atlas of Nearby Galaxies \citep{depaz2007atlas} and the Local Volume Survey \citep{dale2009lvl}. From the 11HUGS sample, the authors selected the galaxies from the 11HUGS sample that had distances of less than 11~Mpc, avoided the Galactic plane ($b\leq 30\arcdeg$), and were bright ($B\leq$ 15~mag). 

\subsection{LEGUS: Legacy Extragalactic UV Survey}
The Legacy Extragalactic UV Survey \citep[LEGUS;][]{calzetti2015legus} used HST to investigate star formation in nearby galaxies and the impact of the environment on formation processes. They observed 50 galaxies within 12~Mpc using 5-band imaging (NUV, U, B, V, I) from the Wide-field Camera 3 (WFC3) and optical imaging with the Advanced Camera for Surveys. The scientific motivation for the survey was to small scale observations (studies of individual star or star cluster on the parsec scale) to larger observations of galaxies (on the kilo-parsec scale) to quantify the clustering of star formation and evolution. The survey obtained star formation histories  from resolved observations of massive stars and extinction-corrected ages and masses of star clusters. The survey sample covered the full range of galaxy properties including mass, morphology, star formation rates, metallicity, and interaction states.

LEGUS attempted to maximize the range of HST observations in order to both resolve and age-date young stellar populations on the parsec scales, and also observe large, kilo parsec scale galaxy structures. This resulted in a final distance range of $\sim$~3.5--12~Mpc which includes the maximum amount of the Local Volume while maintaining quality observations. The survey used a parent sample taken from the 11HUGS catalog \citep{kennicutt2008hugs}. Further cuts were made to the parent sample to maximize UV observation quality, including limiting the minimum distance to 3.5~Mpc and requiring a Galactic latitude $|b| > \ang{20}$ to minimize the effects of foreground extinction. Further, an inclination limit of $i < \ang{70}$ was imposed to minimize dust attenuation along the line of sight. The final survey sample consisted of 50 galaxies. There are additional multi-wavelength observations of a subsample of 260 galaxies from the \textit{Spitzer} Space Telescope IRAC and MIPS, and from the Wide-field Infrared Survey Explorer  mission.

\begin{figure}[h!t]
 \centering
 \includegraphics[width=0.44\textwidth]{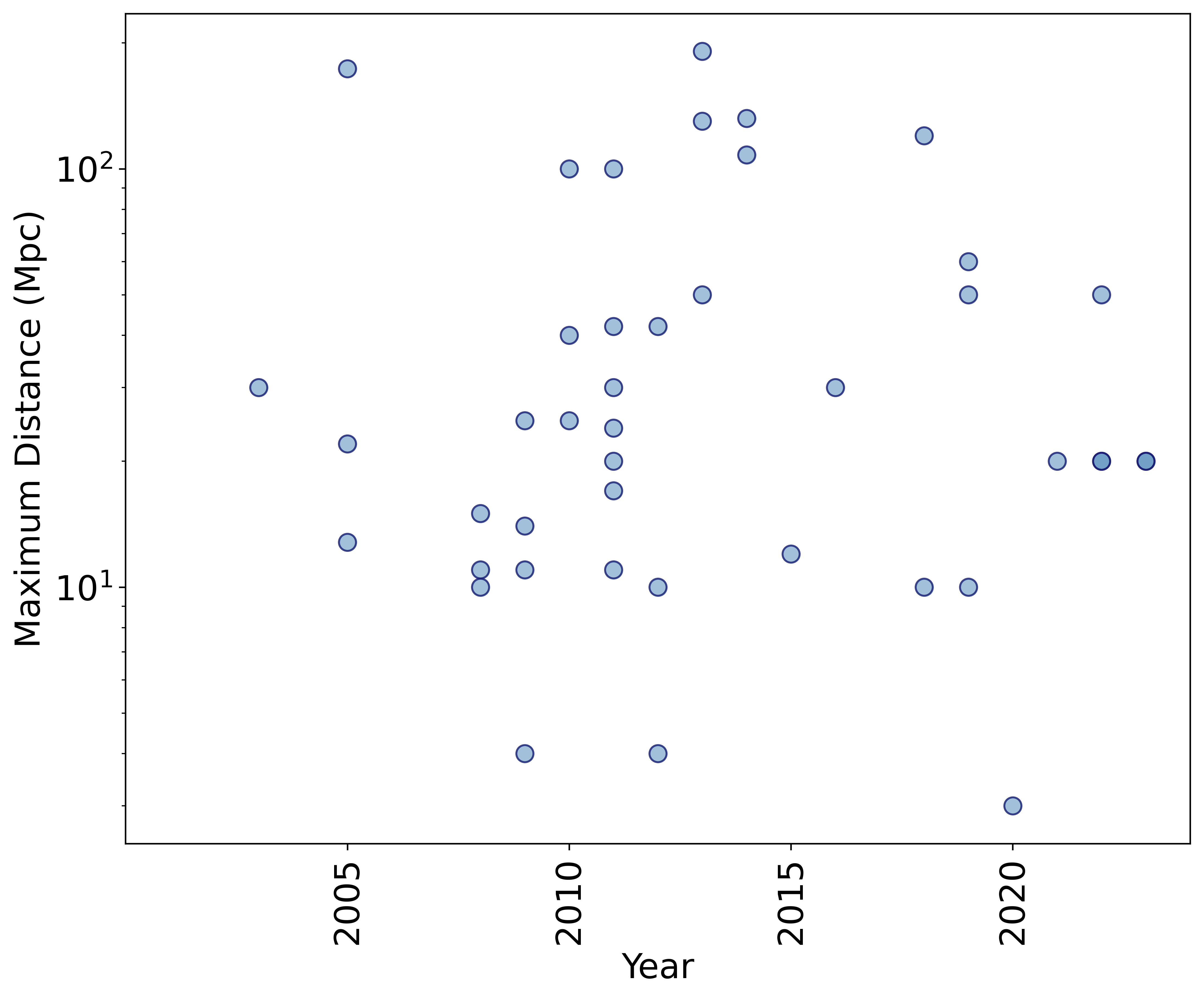}
 \includegraphics[width=0.44\textwidth]{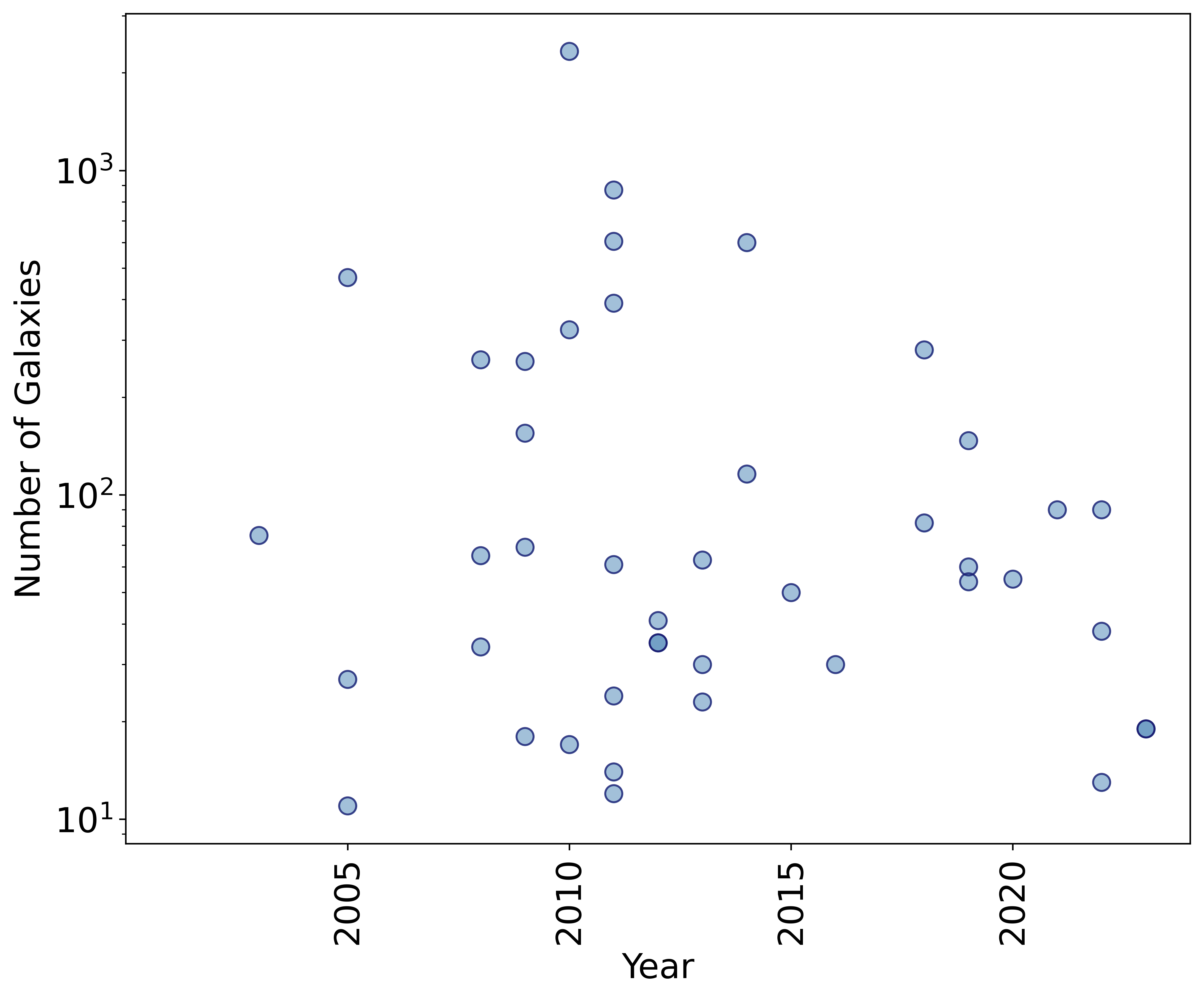}
  \caption{Left: The maximum distance covered by survey samples ordered by publication date. Right: The total number of galaxies included in survey samples ordered by publication date.}
  \label{fig:3}
  \end{figure}

\section{X-ray Surveys}\label{sec: xray}
The main sources of X-ray emission from nearby galaxies are point sources, such as X-ray binaries and supernova remnants, diffuse hot gas, and active galactic nuclei. Sources in the first two categories are, in general, much fainter than AGN. These low luminosities, combined with the limited amount of observing time available on the small number of X-ray telescopes, mean that pointed surveys of nearby galaxies are much less common at X-ray wavelengths compared to other regimes.

\subsection{A \textit{Chandra} Survey of Nearby Spiral Galaxies}
\citet{kilgard2005chandra} carried out a survey of 11 nearby, face-on spiral galaxies with the \emph{Chandra} X-ray Observatory. Unlike most of the other surveys discussed in this work, this survey did not have an official acronym; here we refer to it as ‘CHANDRA’. The goals of the survey were to catalog the populations of X-ray point sources in nearby galaxies, determine their nature and luminosity functions, and to search for variations in these quantities as a function of galaxy properties. The sample was derived from the Nearby Galaxies Catalog \citep{tully1988nearby} and contains galaxies with a range of Hubble types, inclinations $i<35^{\circ}$, distances $\lesssim10$~Mpc, and observed through low Galactic foreground (hydrogen column density $N_{\rm H} < 5\times 10^{20}~$cm$^{-2}$). Observations were carried out with the Advanced CCD Imaging Spectrometer (ACIS) instrument on \emph{Chandra}, with both short and long observations for each galaxy. Exposure times ranged from 2.3 to 98~ks with typical values of about 40~ks. The authors also obtained new optical imaging of the 11 galaxies in broadband $UBVRI$ and narrowband H$\alpha$ and $S_{ii}$ filters. 

\subsection{X-ray spectral survey with XMM-Newton of a complete sample of nearby Seyfert galaxies}
The X-ray Spectral Survey \citep{cappi2006XMM} used X-ray observations to study a sample of 27 Seyfert galaxies in the Local Universe. Using the EPIC CCDs on the XMM-Newton telescope, the survey aimed to test the standard unified model of AGNs on low luminosity galaxies. While previous works had been unable to test the unified model on the low luminosity population, preliminary X-ray studies with new observing technology, such as \citet{ho2001} and \citet{terashima2003}, had opened the door to the possibility that the current model may not hold for less luminous objects. This initial study was a part of a larger project which aimed to characterize and understand the multi-wavelength properties of all 60 Seyfert classifications from the Palomar optical spectroscopic survey of nearby galaxies \citep{panessa2006}. Using the parent sample from the Palomar survey offered the opportunity to perform a multi-wavelength analysis of the Seyfert galaxies \citep{filippenko1985,ho1995}. The original 60 Seyfert galaxies had uniform and high-quality optical data to which additional X-ray observations from the XMM-Newton telescope were added for a smaller select sample. For the purposes of X-ray observations, the 27 nearest Seyfert galaxies with AGNs from \citet{ho1997} were selected, resulting in a distance cut of 22~Mpc. The final sample consisted of 27 Seyfert galaxies, 9 of which were classified as Type 1, and the remaining 18 as Type 2.

\subsection{The \textit{Chandra} Survey of Nearby Galaxies}
The \emph{Chandra} Survey of Nearby Galaxies \citep{sheChandra2017} is the most recent, large-scale collection of X-ray data selected using specific criteria. While this survey does not meet our definition due to the fact that the data were compiled from the \emph{Chandra} Public Archive, it is an important contribution to the extragalactic X-ray regime. The project was focused on the study of active galactic nuclei (AGN) in a total of 719 nearby galaxies. The goal was  to study the relationship between AGN activity and the Hubble type of galaxies, with a special emphasis on low-mass and late-type galaxies. This survey utilized the publicly available ACIS observations of galaxies with a maximum distance of 50~Mpc. 

\section{Results and Discussion} \label{sec: Discussion}
\subsection{Results}
\begin{figure}[h!t]
 \centering
 \includegraphics[width=\textwidth]{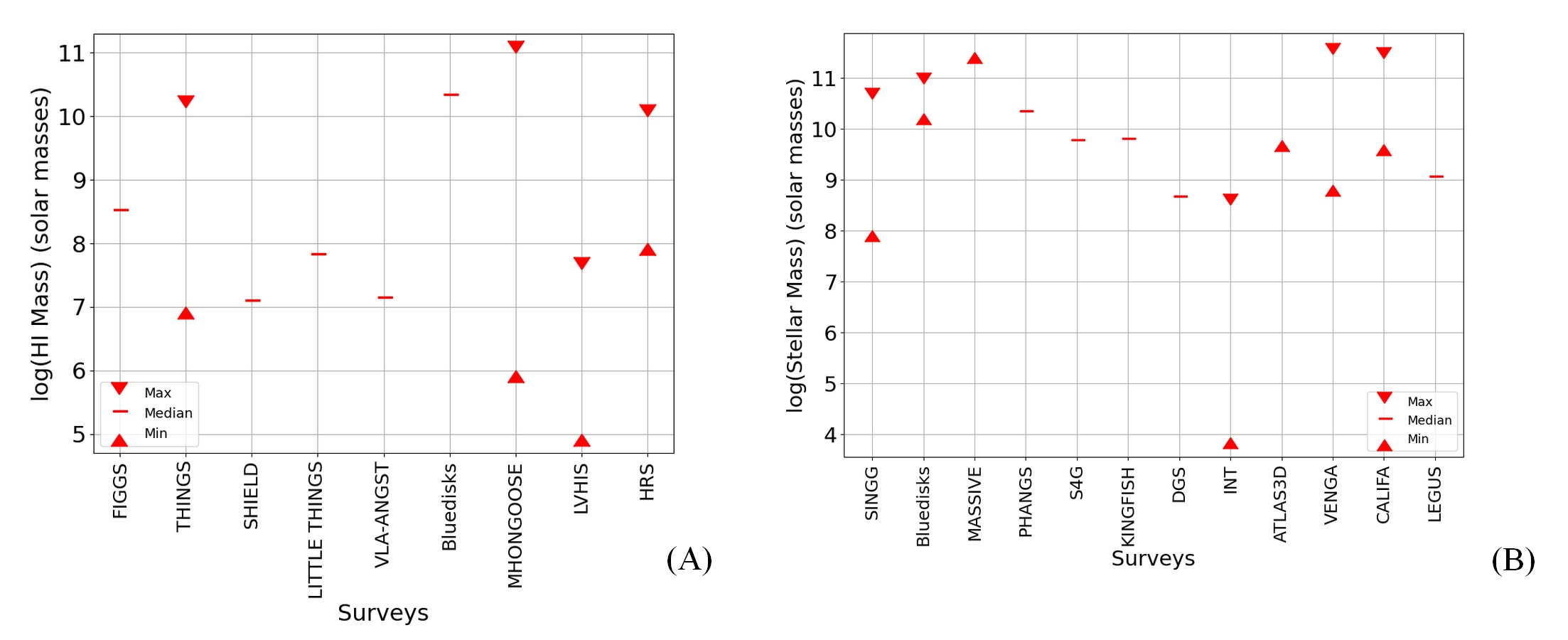}
  \caption{Left: Minimum, maximum and median HI masses for surveys grouped by wavelength, and then by year. Right: Minimum, maximum and median stellar masses for surveys grouped by wavelength, and then by year.}
  \label{fig:4}
  \end{figure}

We observe:

\begin{enumerate}
    \item Maximum Distance: Figure~\ref{fig:3}  shows that the maximum distances reached by nearby galaxy surveys display  significant  variation. The nearest maximum distance  was 4 Mpc (ANGST), however, some surveys pushed past the 100 Mpc mark (e.g. CALIFA, 132 Mpc or SINGG, 174 Mpc). We found that the median of maximum distance values for our sample of surveys was 22 Mpc.
    
    \item Number of Galaxies: there is a wide range in the number of galaxies included within a survey sample, from 11 galaxies (the Chandra Survey of Nearby Galaxies) to over 2000 (S$\mathrm{^4}$G). The median number of galaxies included in samples was 60. We additionally note a slight decrease with time in the number of galaxies included in samples as survey goals become focused on more specific features or processes. This is further complimented by the rise of large catalogs of data derived from recent panoramic surveys such as the Siena Galaxy Atlas \citep{sga2023} or DustPedia \citep{davies2017}. 

    \item Survey Coverage: we note from Figure \ref{fig:1} an abundance of both radio and optical surveys that span from large, statistical surveys to very focused surveys over a large time span. Conversely, the ranges covered by the UV and X-ray regimes are much more limited. UV surveys contain large numbers of galaxies but are concentrated in small distance ranges, while X-ray surveys are incredibly specific in their targets and are limited to intermediate ranges. Both of these regimes have large, publicly available data archives through GALEX and Chandra respectively which cover much larger parameter spaces. These regimes are most affected by observational costs and therefore benefit most from archival data use. As technology continues to evolve and barriers to observations are reduced, these areas are likely to experience the greatest impact and growth. 
    
    \item Mass Range: Figure~\ref{fig:4} shows that there is little variation in the mass ranges covered in either HI mass or stellar mass. The mix of minimum and maximum ranges with median values makes it difficult to make a global statement, however, it appears that there are many surveys with overlapping mass ranges. The main area of overlap is focused on the typical galaxy mass range and there is a clear deficit of surveys at the extreme ends of the mass spectrum. The Massive Survey \citep{ma2014}, a survey focused on targeting only massive galaxies, has the highest median stellar mass. On the other end of the spectrum, the Isaac Newton Telescope Local Group Dwarf Galaxy survey which targets only Local Group dwarfs covers the low-mass range. Most other surreys lie in the intermediate mass range.
\end{enumerate}

\subsection{The evolution of galaxy surveys in the modern era}

The development of new observing technology in the 21st century has resulted in many new galaxy surveys.  Historically, catalogs and atlases of galaxies were focused on compiling basic information about their targets: sky location, morphology, environment, etc.  In the early turn of the century, we saw the rise of imaging-based surveys. These images were used for simple measurements and analyses, though the main data outputs were often the images themselves. Recently, however, surveys have focused on computing and compiling detailed measurements for specific properties and processes within galaxies, consistent with their very specific science goals and selection criteria. These surveys identify their targets using imaging and classification data from previous atlases and legacy surveys, or from large all-sky surveys which can provide a better parent sample to search for galaxies with atypical properties.

As the focus of galaxy surveys continues to evolve, so does the target selection process. For the legacy observing projects, the majority of selection criteria was determined by instrumental limitations. Criteria were set to ensure quality observations and a time-efficient project. Survey architects typically selected objects that would be easy to observe, which preferred bright galaxies close enough to achieve high spatial resolution. This was often the determining factor in the definition of ``nearby".
More recent surveys greatly benefited from previous large, legacy observing programs that provided their parent samples. As surveys became more focused on specific features of galaxies or processes, the selection criteria also became more specific. Galaxy orientation and location were taken into consideration to ensure that the feature(s) of interest could actually be observed. For these programs, the definition of nearby was also influenced by the number of potential targets within different distances, balancing the need for high resolution observations with the desire to create a statistically significant sample.

 \begin{figure}[h!t]
 \centering
 \includegraphics[width=0.9\textwidth]{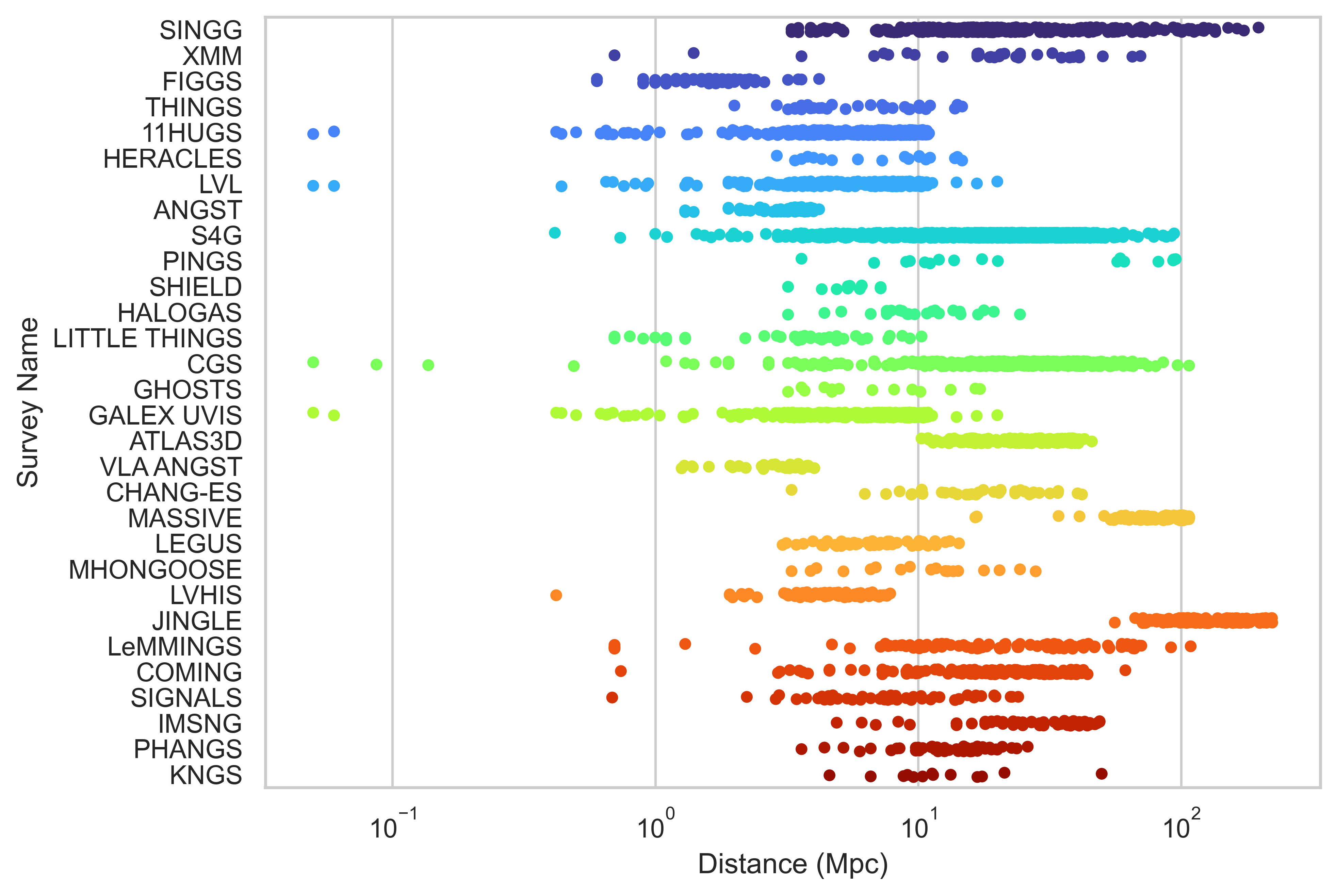}
  \caption{The distribution of distances to targets included in the surveys from this paper. The surveys are ordered chronologically and a slight overall increase in distance can be observed over time. Not shown are NGLS, a Chandra Survey of Nearby Spiral Galaxies, and the Isaac Newton Telescope Monitoring Survey, for which individual galaxy distances were not tabulated in the survey description.}
  \label{fig:5}
  \end{figure}

As observational technology and scientific interests evolved,  multi-wavelength observations of galaxies gained additional importance. The larger legacy surveys such as SINGS \citep{kennicutt2003sings} or the \textit{Herschel} Reference Survey \citep{boselli2010herschel} were often adopted as parent samples. In some cases, observations from these surveys were not sufficient to fulfill the science goals of a new survey, particularly for studies of very specific parameters or processes. In such cases, galaxies might be cross-matched between multiple all-sky surveys to piece together additional observations. A few surveys have skipped this step of piecing together surveys by running simultaneous observational projects in multiple wavelengths. The most recent large-scale example of this type of project is the Physics at High Angular resolution Nearby Galaxies Survey (PHANGS) where the initial survey, PHANGS-ALMA, provided the parent sample for the entire project. The following four sub-surveys (HST, MUSE, H-$\mathrm{\alpha}$, and JWST) followed a top-down approach in which each subsequent survey was a subset of the previous surveys. This ensured that, as science goals became progressively more specific, there were multi-wavelength observations for a sample of galaxies. 

In parallel to the rise of focused surveys, a new cohort of large-scale galaxy projects have gained popularity. The early 2000s saw the launch of many new space telescopes which revolutionized the way that astronomers surveyed the sky. The Sloan Digital Sky Survey spent the first eight years of operation (from 2000 to 2008) creating an deep, multi-wavelength imaging atlas of the sky \citep{SDSS2003}. Its first data release coincided with the launch of the Galaxy Evolution Explorer \citep[GALEX;][]{martin2003} and the Spitzer Space Telescope \citep{werner2004} in 2003. These massive panoramic efforts allowed for new science investigations to be performed on existing data from the archives at a previously unprecedented rate. These surveys removed the observational cost of galaxy surveys while maintaining the necessary depth and resolution required for detailed measurements. As a result, there has been a recent surge of new catalogs that include unprecedented numbers of galaxies with detailed measurements. Similarly, the WISE Extended Catalog used WISE IR observations in order to measure the global photometric properties of the largest 100 nearby galaxies \citep{jarrett2019WISE}, while DustPedia combined archival \textit{Herschel} and \textit{Planck} data for nearly 1000 galaxies in order to study dust in the local Universe \citep{davies2017}. These catalogs compile detailed measurements for a sample size that had not before been seen. On an even larger scale, the Siena Galaxy Atlas contains combined optical imaging from the Dark Energy Survey and infrared WISE observations in order to produce photometry data, precise coordinates, and surface brightness profiles for a sample of over 20 000 galaxies \citep{moustakas2023SGA}. New and upcoming space missions, such as \textit{Euclid} and the \textit{Nancy Grace Roman Space Telescope}, will do the same for space-based panoramic optical imaging in the near future.
\begin{figure*}[h!t]
 \centering
 \includegraphics[width=0.9\textwidth]{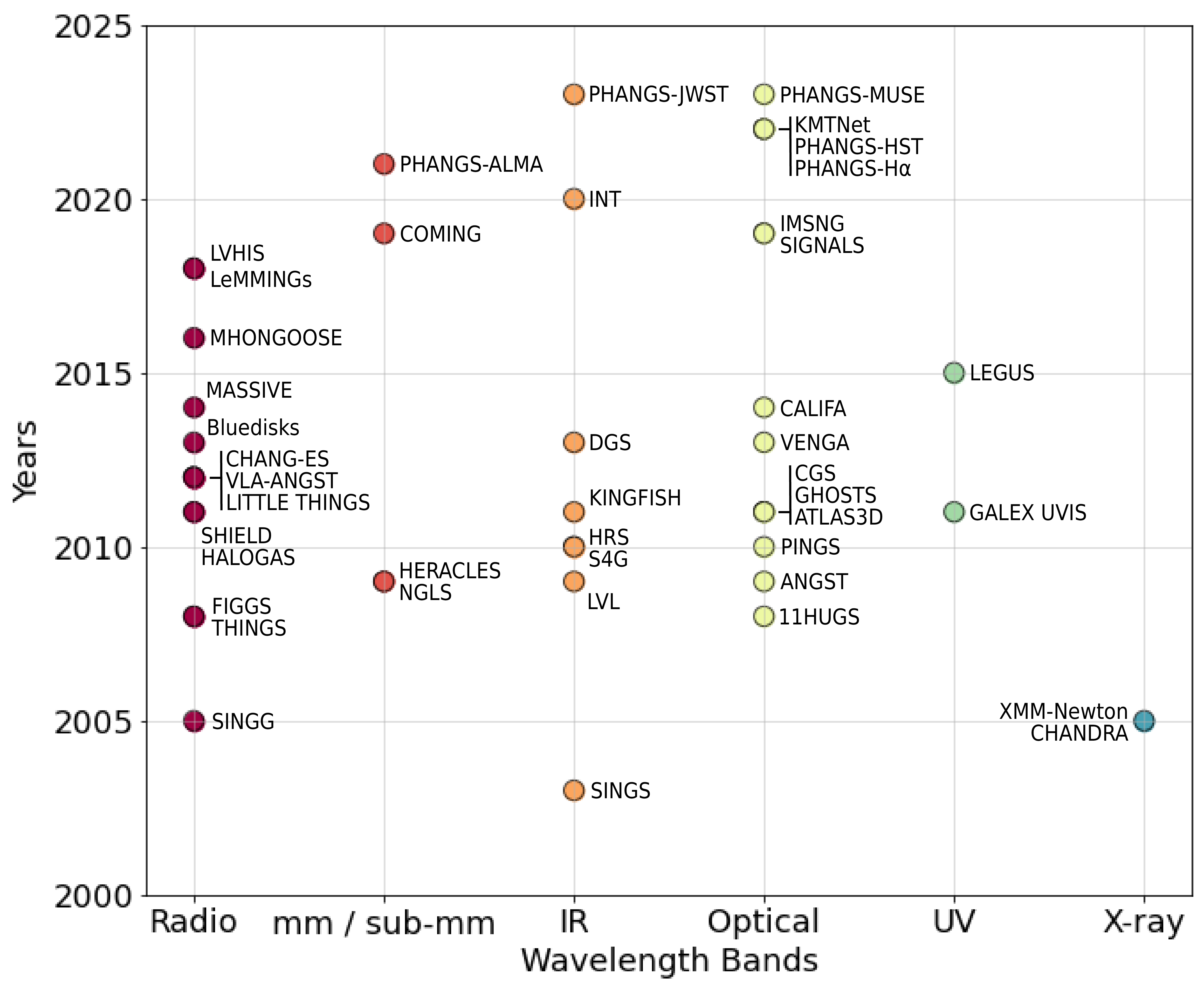}{}
  \caption{The coverage of different bands by nearby galaxy surveys ordered by year. Each colored column represents a different wavelength domain, from radio to X-ray. The optical and radio regimes are more densely populated, while mm/sub-mm, UV and X-ray surveys are less common.}
  \label{fig:6}
  \end{figure*}
\subsection{Definition of \textit{nearby} in galaxy surveys}
With the changes that galaxy surveys have undergone in the past two decades, it's not surprising that the parameters covered in surveys have also changed. One parameter which has experienced a large change is the distance range covered (Figure~\ref{fig:5}). Through the review of the surveys in this paper, we find that the term \textit{nearby} does not have a widely accepted value when used to refer to the distances to surveyed galaxies. Its use ranges from describing galaxies within only the Local Group \citep[ANGST;][]{dalcanton2009ANGST}, to the collection of galaxies within the Local Volume \citep[LVL;][]{dale2009lvl}, to galaxies at distances stretching out to nearly 200 Mpc \citep[SINGG;][]{meurer2006singg}. Even terms such as ``nearby universe" or ``Local Volume" often refer to varying distances. We also see an increase in the maximum distance as time progresses.  

We further see that surveys attempt to include statistically complete samples. Surveys with goals of completeness in different physical parameters (morphology, environments, etc) will have their distance limit set by the locations of the targets required to reach completeness. As studies transition from studying typical populations of nearby galaxies to less common galaxy types, we see a distinct decline in the number of galaxies within surveys while the distance ranges remain similar. This effect is especially prominent in surveys targeting very specific processes or features, such as KMTNet \citep{byun2022kmtnet}, which targets only spiral galaxies with low-surface brightness features. 

\subsection{How complete are the observational archives for nearby galaxies?}
In this work, we have summarized a total of 43 pointed nearby galaxy surveys dating back to 2003. These large-scale, legacy surveys have created an excellent base to build a general model for galaxy evolution. Many surveys have observed ``normal" galaxies (in terms of color, star formation rates, and gas density) with a particularly high number of optical and radio surveys as seen in Figure \ref{fig:6}. On the other hand, there is a lack of short wavelength studies (in UV and X-rays), limiting extragalactic studies of high-mass stars, the hottest phases of other galaxies' interstellar media, and their compact object systems. However, without observations of galaxies that fall outside normal ranges, a complete picture of galaxy evolution in the nearby universe will remain impossible.

Apart from selection criteria, there remain limitations to the targets that can be selected for observation. For example, galaxies located along the Galactic plane suffer from much higher extinction and are more difficult to efficiently observe. Atmospheric absorption means that some wavelengths must be observed from space. The costs associated with space-based observing missions are much greater than with ground-based programs which affects the number of observations that can be taken. Most notably, there are fewer pointed galaxy surveys within the X-ray regime due to the highly expensive nature of X-ray observations. 
To compensate for this, large public archives of data have been compiled, such as the \emph{Chandra} Survey of Nearby Galaxies \citep{sheChandra2017}. Observational costs also affect the number of synoptic surveys that are available. Programs which involve monitoring targets over extended time periods is extremely resource intensive. Although public, synoptic surveys of nearby galaxies have to date been limited, this situation will soon change drastically with the imminent optical Legacy Survey of Space and Time on the Vera C. Rubin Observatory and radio surveys with the SKA Observatory.

Finally, some populations of galaxies are simply more difficult to observe due to their size (e.g. dwarf galaxies) or brightness (low-surface brightness galaxies). As technology and observing techniques continue to advance, the effects of these limitations will undoubtedly decrease. 

\section{Conclusions}
There has been immense growth and change in the field of astronomy since the beginning of the 21st century, and nearby galaxy surveys are no exception. Within the 43 surveys included in this paper, we note a significant evolution in the types of programs, the motivating science questions, and the target selection process. The archives of observational data are well populated and cover the range from extensive general observations to detailed, high resolution observations. There is also a clear advantage in having multi-wavelength coverage to obtain complete understanding of the processes that drive galaxy evolution. This can occur as new surveys are initiated from larger parent samples from well-known legacy surveys such as S$\mathrm{^4}$G \citep{sheth2010spitzer}, NGLS \citep{wilson2009james}, or LVL \citep{dale2009lvl} or from a larger program perspective such as PHANGS \citep{leroy2021phangs}. Further, observing technology is rapidly reaching the threshold in which large, panoramic observing programs are able to obtain comparable observations, in terms of quality and depth, as traditional targeted surveys. These new archives of data have further assisted in the data explosion the field has recently experienced. Finally, while there is coverage from nearby galaxy surveys in all observational bands, areas such as UV and X-ray are the least populated by galaxy surveys and are likely to experience the most growth due to technology evolution in the future.

\begin{acknowledgments}
We acknowledge that Western University is located on the traditional lands of the Anishinaabek, Haudenosaunee, Lūnaapéewak and Chonnonton Nations, on lands connected with the London Township and Sombra Treaties of 1796 and the Dish with One Spoon Covenant Wampum. This land continues to be home to diverse Indigenous Peoples (First Nations, Métis and Inuit) whom we recognize as contemporary stewards of the land and vital contributors of our society. 

We further acknowledge historical and ongoing injustices that Indigenous Peoples (First Nations, Métis and Inuit) endure in Canada, and we, as researchers at a public institution, consider it our responsibility to contribute toward revealing and correcting miseducation as well as renewing respectful relationships with Indigenous communities through our research.

The authors gratefully acknowledge support from a Discovery Grant from the Natural Sciences and Engineering Research Council of Canada to PB. We thank the anonymous referee for the suggestions which helped improve this paper. HSC thanks the support of Sarah Gallagher through numerous insightful conversations. 
\end{acknowledgments}  

\software{astropy \citep{2013astropy}}

\bibliography{ref}{}
\bibliographystyle{aasjournal}

\end{document}